\documentclass[aps,prb,reprint,groupedaddress,amsmath,amssymb,longbibliography]{revtex4-2}

\usepackage{graphicx} 
\usepackage{dcolumn} 
\usepackage{bm}
\usepackage[pdfstartview=FitH, CJKbookmarks=true, bookmarksnumbered=true, bookmarksopen=true, colorlinks=true, pdfborder=0 0 1, citecolor=blue, linkcolor=blue, linktocpage=true] {hyperref}
\usepackage{epstopdf} 

\begin{document}

\title{Topological invariants for interacting systems: from twisted boundary condition to center-of-mass momentum}

\author{Ling Lin$^{1,2,3}$}

\author{Yongguan Ke$^{2,3}$}
\email{Email: keyg@mail2.sysu.edu.cn}

\author{Chaohong Lee$^{1,2,3}$}
\email{Email: chleecn@szu.edu.cn, chleecn@gmail.com}

\affiliation{$^{1}$College of Physics and Optoelectronic Engineering, Shenzhen University, Shenzhen 518060, China}
\affiliation{$^{2}$Guangdong Provincial Key Laboratory of Quantum Metrology and Sensing $\&$ School of Physics and Astronomy, Sun Yat-Sen University (Zhuhai Campus), Zhuhai 519082, China}
\affiliation{$^{3}$State Key Laboratory of Optoelectronic Materials and Technologies, Sun Yat-Sen University (Guangzhou Campus), Guangzhou 510275, China}

\date{\today}

\begin{abstract}
Beyond the well-known topological band theory for single-particle systems, it is a great challenge to characterize the topological nature of interacting multi-particle quantum systems.
Here, we uncover the relation between topological invariants defined through the twist boundary condition (TBC) and the center-of-mass (c.m.) momentum state in multi-particle systems.
We find that the Berry phase defined through TBC can be equivalently obtained from the multi-particle Wilson loop formulated by c.m. momentum states.
As the Chern number can be written as the winding of the Berry phase, we consequently prove the equivalence of Chern numbers obtained via TBC and c.m. momentum state approaches.
As a proof-of-principle example, we study topological properties of the Aubry-Andr{\'e}-Harper (AAH) model.
Our numerical results show that the TBC approach and c.m. approach are well consistent with each other for both many-body case and few-body case.  
Our work lays a concrete foundation and provides new insights for exploring multi-particle topological states.  
\end{abstract}
\maketitle

\section{Introduction}

Since the discovery of quantum Hall effect~\citep{PhysRevLett.45.494}, topological quantum states have been widely and intensively studied.
Owing to topological band theory, various quantum topological states have been successfully found in non-interacting systems~\citep{RevModPhys.82.3045, RevModPhys.83.1057, kane2013topological, RevModPhys.88.035005}.
However, in the presence of particle-particle interaction, because the single-particle quasi-momentum is not a good quantum number, topological band theory usually fails.
In interacting many-body quantum systems, different theoretical frameworks are developed to explore fascinating strongly correlated topological phases such as the fractional quantum Hall effect~\citep{PhysRevLett.48.1559, PhysRevLett.50.1395,RevModPhys.71.875}.

The first attempt is to introduce the twisted boundary condition (TBC) to define a topological invariant~\citep{niu1984quantised,PhysRevB.31.3372, RevModPhys.82.1959, PhysRevLett.122.146601,PhysRevLett.80.1800, PhysRevX.8.021065,PhysRevB.103.224208, PhysRevLett.110.075303,PhysRevB.100.054108, PhysRevResearch.2.042024} for interacting many-body quantum systems.
Similar to the periodic boundary condition (PBC), under the TBC, the boundaries along the same direction are glued together.
The essential difference is that particles gain extra phases when they go through the boundaries under the TBC.
The extra phase, which is known as the twist angle, can be considered as a result of inserting magnetic flux~\citep{PhysRevB.23.5632,PhysRevLett.96.060601} whose change will induce the flow of current~\citep{PhysRevLett.68.1375, PhysRevB.106.045410}.
Topological invariants defined via the twist angle have successfully explain topological features related to the system's response to external fields, such as, the polarization (Berry phase)~\citep{PhysRevB.27.6083, PhysRevB.47.1651,PhysRevB.48.4442, RevModPhys.66.899} and the quantized Hall conductance~\citep{PhysRevLett.49.405}.

In recent years, a new approach has been proposed via the co-translation symmetry~\citep{PhysRevB.96.195134,Qin_2018,PhysRevB.96.195134,PhysRevA.95.063630}, 
with which the total energy remains unchanged when all particles as a whole are shifted by unit cells.
The co-translation symmetry supports the c.m. quasi-momentum as a good quantum number, and enables few-body topological band theory, in which topological invariants of gapped few-body Bloch bands can be defined via c.m. quasi-momentum states~\citep{PhysRevB.96.195134,Qin_2018,PhysRevB.96.195134,PhysRevA.95.063630}.
Here, ``few-body" means that the total particle number is fixed as a finite value $N$ (even in the thermodynamic limit $ L\to \infty$).
With this approach, exotic interacting topological phases have been uncovered, such as topological bound edge states~\citep{PhysRevB.96.195134,Qin_2018,PhysRevB.96.195134,PhysRevA.101.023620}, topologically resonant tunnelings~\citep{PhysRevA.95.063630}, and interaction-induced Thouless pumping~\citep{PhysRevA.101.023620}.

\begin{figure}
  \includegraphics[width = \columnwidth ]{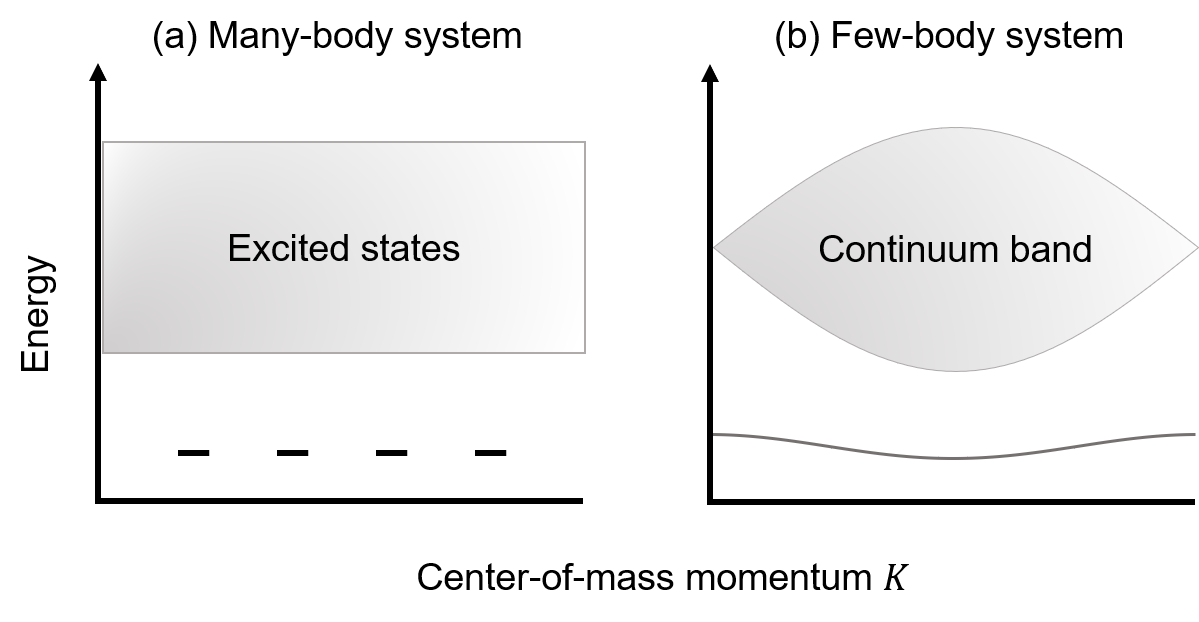}
  \caption{\label{fig:FIG_Spectrum_illustration}
  Illustrative diagram for low-lying energy spectra of (a) many-body systems and (b) few-body systems under the PBC.
  For many-body systems, the degeneracy of the gapped ground-state manifold is finite, which strongly depends on the filling factor $\nu=N/L$.
  Here, we show the four-fold degenerate ground states as an example.
  For few-body systems, there appears band-like structure.
  Bands are dispersive and continuous in the thermodynamic limit.
  When the interaction is strong enough, there appears gapped isolated band emerging from the continuum band, which usually correspond to bound states.
  }
\end{figure}

While there appear extensive interests in few-body topological states~\citep{PhysRevB.96.195134, Qin_2018, PhysRevA.95.063630, PhysRevA.97.013637,PhysRevA.97.013637,PhysRevA.101.023620,Marques_2018,PhysRevResearch.2.013348, PhysRevResearch.2.033267,Mei_2019,Malki_2020,PhysRevA.95.033831, PhysRevResearch.2.033190},
it is more challenging and appealing to study topological states in many-body systems, where the filling factor $\nu=N/L$ keeps a finite value (even in the thermodynamic limit $L\to\infty$).
In fact, the number of gapped many-body ground states strongly depends on the filling factor~\citep{PhysRevLett.84.1535}.
%
%
In contrast to the continuous band structure in few-body systems, many-body gapped ground-state manifold may only consist of finite degenerate eigenstates with certain quasi-momenta; as depicted in Fig.~\ref{fig:FIG_Spectrum_illustration} (a). 
It seems that there is no well-defined band structure for many-body systems.
%
Up to now, how to utilize quasi-momentum states to characterize many-body topological states remains vague.
%

Although the TBC approach and the c.m. momentum approach seem apparently different, they can independently and faithfully define topological invariants for interacting multi-particle systems. 
Up to now, there is no comparison of the physics obtained from applying the two approaches to a same system. 
Understanding the relation between the two approaches can give new insights into the foundation of interacting topological states.
It is already known that threading magnetic field to a system will induce a shift of the c.m. quasi-momentum, indicating that the twisted angle has the same status of the c.m. quasi-momentum.
However, it is unclear whether the topological invariants defined with these two approaches are equivalent.

\begin{figure*}
  \includegraphics[width = 0.85\textwidth ]{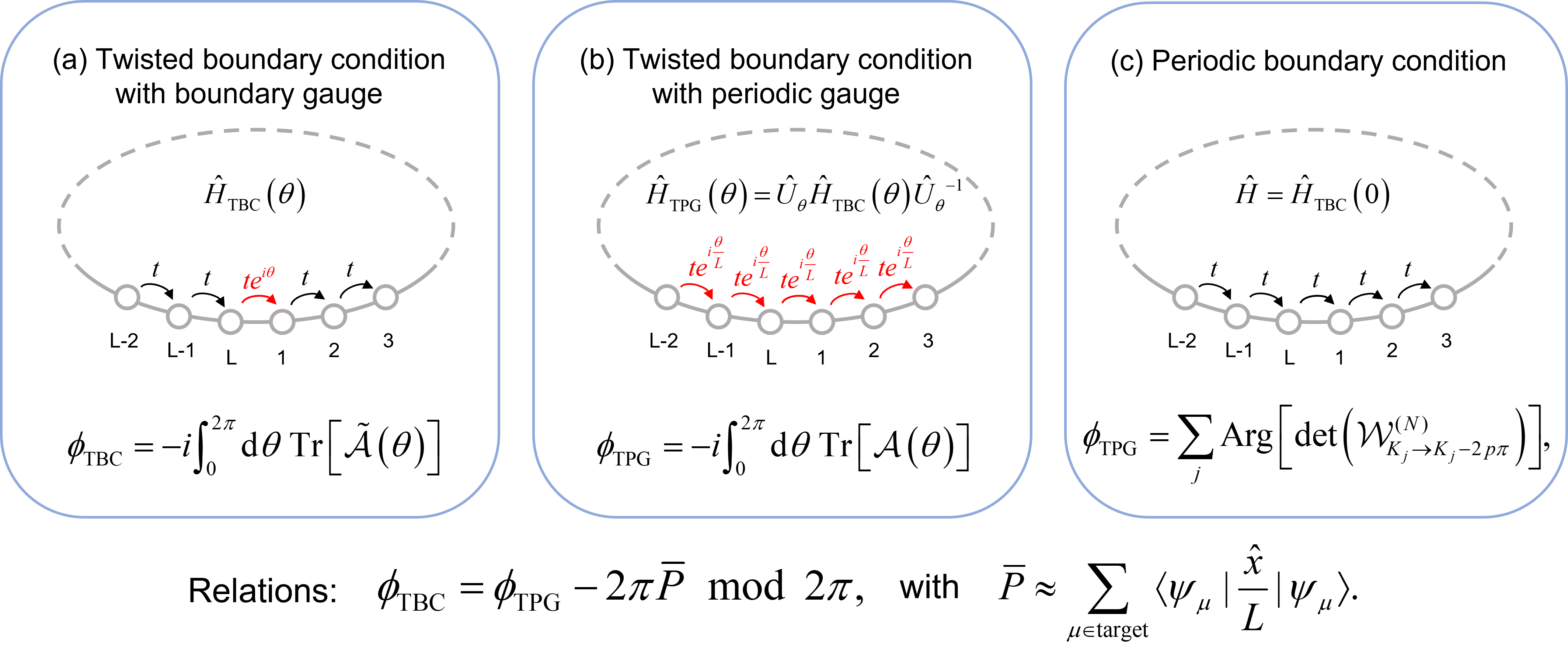}
  \caption{\label{fig:FIG_TBC_Gauge}
  Schematic demonstration of the simple tight-binding 1D lattice under different boundary conditions and their corresponding Berry phases.
  (a) the general TBC with boundary gauge, (b) the TBC with periodic gauge, and (c) the periodic boundary condition.
  Under the boundary gauge (a), a particle only gains the phase when it crosses the boundary.
  Under the periodic gauge (b), a particle gains a ``diluted" and homogeneous phase everywhere during the tunneling.
  The arrows indicates the tunneling of particles.
  Here $t$ is the tunneling strength, $\theta$ is the twist angle, and $L$ is the total number of cells.
  In both (b) and (c), the system possesses the co-translational symmetry.
  The Berry phases corresponding to these three configurations as well as their relations are given below.
  The central result is that the TBC Berry phase can be formulated by the c.m. momentum states.
  }
\end{figure*}

In this work, we generalize the c.m. momentum approach to many-body systems by introducing the \emph{multi-particle Wilson loop}, and systematically clarify the relation between the TBC approach and the c.m. momentum approach; as depicted in Fig.~\ref{fig:FIG_TBC_Gauge}.
Under TBC, we classify two different gauges as: (i) boundary gauge in which the twist angle is only gained at crossing boundary, and (ii) periodic gauge in which twist angle is uniformly distributed at each hopping term.
With periodic gauge under TBC, the co-translation symmetry is restored, and the c.m. momentum is related to the twist angle.
The Berry phases defined via the twist angle under boundary and periodic gauges only differ by a trivial classical polarization, dubbed the TBC Berry phase for brevity.
Under PBC, by introducing the multi-particle Wilson loop, the Berry phase can be obtained from the c.m. momentum states, dubbed the c.m. Berry phase for brevity.
The multi-particle Wilson loop is a generalization from the single-particle Wilson loop, applicable to both few-body to many-body systems.  
By employing perturbative analysis, we uncover that the TBC Berry phase in periodic gauge can be equivalently obtained via c.m. quasimomentum states and is related to the c.m. Berry phase.
Since the Chern number can be expressed as the winding of Berry phases in two-dimensional (2D) systems, the Chern numbers obtained via the TBC approach and the c.m. momentum approach are therefore equivalent.
To verify our general arguments, we consider a Aubry-Andr{\'e}-Harper (AAH) model and numerically compute the topological properties of the gapped state.
Our results clearly show that the two Berry phases as well as the Chern numbers defined through the twist angle and the c.m. momentum state are consistent with each other in both the few-body and many-body cases.

The rest of this article is organized as follows. 
In Sec.~\ref{Sec:TBC_and_CM_momentum}, we introduce and review some key properties of the twisted boundary condition and the co-translation symmetry.
We then discuss the relation between the twist angle and the center-of-mass momentum.
In Sec.~\ref{Sec:Zak_phase}, we discuss the Berry phase and the Chern number defined through the twist angle and the c.m. momentum state.
We derive the relation between the TBC Berry phase and the c.m. Berry phase by using perturbative expansion, and we discuss it respectively for many-body and few-body systems.
In Sec.~\ref{Sec:Numerical_verification}, we illustrate our general framework through the AAH model numerically and verify our arguments.
In Sec.~\ref{Sec:Summary}, we briefly summarize and discuss our results.

\section{Twisted boundary condition and center-of-mass momentum}
\label{Sec:TBC_and_CM_momentum}

In this section, we focus on discussing the relation between the twist angle under TBC and the c.m. momentum state under PBC. 
 We will review the concepts of the twisted boundary condition, introduce the co-translation symmetry and the c.m. momentum, and then show the relation between the twist angle and the c.m. momentum state for both few-body and many-body systems.

\subsection{Twisted boundary condition}
\label{Sec:sub_TBC}

To illustrate the TBC, we consider a generic form of the one-dimensional (1D) Hubbard-like Hamiltonian with two-body interaction
\begin{eqnarray}
\label{eqn:Ham_TBC_theta}
 &&{\hat{H}_{\text{TBC}}}\left( {\theta } \right)=-\sum\limits_{{x},{d}}{\left( {{t}_{{d}}}{{e}^{i {\Theta _{{x}+{d},{x}}}}}\hat{c}_{{x}+{d}}^{\dagger }{{{\hat{c}}}_{{x}}}+H.c. \right)}  \nonumber \\
&& \qquad \qquad \qquad+\sum\limits_{{x},{d}}{{{V}_{{d}}}{{{\hat{n}}}_{{x}+{d}}}{{{\hat{n}}}_{{x}}}};  \\
 &&{\Theta _{{{x}} + {{d}},{{x}}}} = \left\{ {\begin{array}{*{20}{c}}
{{\theta},}&{\;\;\langle {{x}} + {{d}},{{x}}\rangle \; \mathrm{cross}\;\mathrm{the}\;\mathrm{boundary,}}\\ 
{0,}&{{\mathrm{otherwise}}.}
\end{array}} \right. \nonumber
\end{eqnarray}
in which $\hat{c}_{{x}}^{\dagger }\;\left( {{{\hat{c}}}_{{x}}} \right)$ is the creation (annihilation) operator at the ${x}$-th site, $t_{{d}}$ and $V_{{d}}$ are the tunneling strength and the two-body interaction strength, respectively.
Tunneling and interaction are both finite-range and only dependent on the relative distance ${d}$, which ensures the co-translation symmetry when $ \theta = 0$.
For simplicity, the lattice constant (i.e. the distance between two neighboring lattice sites) is set as ${a}=1$.
Meanwhile, the boundary is positioned between the $L$-th and $1$-st cells, and particles will gain phase only when tunneling through this boundary.
Hence, the TBC can be viewed as a generalized periodic boundary condition, see Fig.~\ref{fig:FIG_TBC_Gauge} (a) for schematic demonstration.
It can be noted that the Hamiltonian \eqref{eqn:Ham_TBC_theta} is a periodic function of ${\theta}$ with the period $2\pi$, that is,
\begin{equation}
{{{\hat{H}}}_{\text{TBC}}}\left( {\theta }+2\pi \right)={{{\hat{H}}}_{\text{TBC}}}\left( {\theta } \right).
\end{equation}

The twist angle $\theta$ can be seen as a consequence of the insertion of a magnetic field, and particles feel gauge field in the lattice.
Due to the gauge freedom, there are numerous conventions to determine how the vector potential (gauge field) distributes.
Here, we choose a particular gauge by introducing the twist operator \citep{PhysRevB.98.155137},
\begin{equation}
\label{eqn:U_theta_operator}
{{{\hat{U}}}_{{\theta }}}=\exp \left( i{{\frac{\theta }{L}}} \hat{{x}} \right),
\end{equation}
in which $\hat{x}=\sum\nolimits_{{x}}{{x}{{{\hat{n}}}_{{x}}}}$ is the position operator and $L$ is length of the system.
It can be checked that ${{{\hat{U}}}_{{\theta }}}\hat{c}_{{x}}^{\dagger }{{{\hat{U}}}_{{\theta }}}^{-1}={{e}^{i {{\frac{\theta }{L}}}{ x }}}\hat{c}_{{x}}^{\dagger }$.
Under this twist transformation, the interaction type considered here remains unchanged.
Then, the unitarily equivalent Hamiltonian reads as
\begin{eqnarray}
\label{eqn:Ham_PBC_theta}
  {{{\hat{H}}}_{\text{TPG}}}\left( {\theta } \right) &=& {{{\hat{U}}}_{{\theta }}}{{{\hat{H}}}_{\text{TBC}}}\left( {\theta } \right){{{\hat{U}}}_{{\theta }}}^{-1} \nonumber \\
&=& -\sum\limits_{{x},{d}}{\left( {{t}_{{d}}}{{e}^{i{{\frac{\theta }{L}}} {d}}}\hat{c}_{{x}+{d}}^{\dagger }{{{\hat{c}}}_{{x}}}+H.c. \right)} \nonumber \\
&& +\sum\limits_{{x},{d}}{{{V}_{{d}}}{{{\hat{n}}}_{{x}+{d}}}{{{\hat{n}}}_{{x}}}}.
\end{eqnarray}
The above transformation is generally called the large gauge transformation \citep{PhysRevLett.84.3370}.
This particular unitary transformation means that we have chosen a gauge such that the vector field distributes uniformly, which is beneficial for us to establish the relation between the c.m. momentum and the twist angle later.
For convenience, we refer to the gauge choice in the transformed Hamiltonian \eqref{eqn:Ham_PBC_theta} as the \emph{periodic gauge} [see Fig.~\ref{fig:FIG_TBC_Gauge} (b)], since the system satisfies the periodic boundary condition (PBC).
Notably, the periodicity of the Hamiltonian with respect to the twist angle $\theta$ is no longer $2\pi$ under this gauge.
For the twist angle appearing only at the boundary [see Hamiltonian~\eqref{eqn:Ham_TBC_theta}], we call it \emph{boundary gauge} [see Fig.~\ref{fig:FIG_TBC_Gauge} (a)].
In Fig.~\ref{fig:FIG_TBC_Gauge}, we give a simple demonstration to show the essential differences among the TBC with the boundary gauge, the TBC with the periodic gauge, and the PBC without the twist angle.
We would like to stress that the energy spectrum under TBC is independent of the gauge choice.
%

\subsection{Co-translation symmetry and center-of-mass momentum}
\label{Sec:sub_Cotranslation_Sym_and_CM_momentum}

In a general interacting multi-particle system, the single-particle translation symmetry is broken.
However, the periodic system remains invariant after the translation of all particles when the inter-particle interaction depends only on their relative distance.
The translation of all $N$ particles in a 1D lattice \cite{PhysRevB.96.195134} can be expressed as 
\begin{equation}
\hat T|{x_1},{x_2}, \cdots ,{x_N}\rangle  = |{x_1} +1,{x_2} + 1, \cdots ,{x_N} + 1\rangle ,
\end{equation} 
where $\hat{T}$ is the co-translation operator translating \emph{all} particles for a unit cell, $|{x_1},{x_2}, \cdots ,{x_N}\rangle$ is the multi-particle basis in position space, with ${x}_j$ referring to the position of the $j$-th particle.
In our analysis, we assume the particles to be bosonic.
For fermions, although the translation is similar, one should take care of the periodic boundary condition and the anti-commutation relation, see Appendix.~\ref{appendix:co_translation_fermions} for detailed discussions.

Alternatively, we can use a c.m. position basis to expand the $N$-particle states,
\begin{equation} 
|{R},\beta \rangle \Leftrightarrow  |{x_1},{x_2}, \cdots ,{x_N}\rangle, 
\end{equation}
where ${R}=\sum\nolimits_{j}{{{{x}}_{j}}}/N$ is the c.m. position of the $N$ particles, $\beta$ is an \emph{abstract label} corresponding to the relative distribution of the $N$ particles.
Mathematically, $\beta$ is uniquely determined by the set of all relative positions: $\left\{ {{{x}}_{i,j}} \right\}_\beta$, ${x}_{i,j}={x}_i-{x}_j,\; 1 \le i < j \le N$.
That is, $\left\{ {{{x}}_{i,j}} \right\}_\beta$ contains the information of all relative positions between any pairs of particles.
States labeled with the same $\beta$ share the same relative distribution and can be translated to each other via the co-translation operation.
In other words, these states form an invariant subspace for the representation of co-translation operator.
The number of $\beta$ depends on the geometry of the lattice and the total number of particles, which grows rapidly with the system size and the total number of particles.

Since the co-translation operation will not change the relative positions between particles, we have
\begin{equation}
\hat{T}|{R},\beta \rangle =|{R}+{1},\beta \rangle.
\end{equation}
%
%
The co-translation symmetry is defined by the commutation between the co-translation operator and the Hamiltonian: $[ { \hat{H}, \hat{T}} ] = 0$.
If a multi-particle system under PBC has co-translation symmetry, although the co-translation symmetry is broken under TBC with boundary gauge, it can be restored under TBC with periodic gauge.

Eigenstates of the co-translation operator can be expressed as
\begin{equation}
\label{eqn:Translational_Operator_Eigenstates}
|{K},\beta \rangle =\frac{1}{\sqrt{C_\beta}}\sum\limits_{{R}}{{{e}^{i{K} {R}}}|{R},\beta \rangle }
\end{equation}
where ${K}$ is a good quantum number and $C_\beta$ is the normalization factor. 
The summation in Eq.~\eqref{eqn:Translational_Operator_Eigenstates} is over all multi-particle position basis having the same relative distribution characterized by $\beta$, and the normalization factor depends on the number of these position basis.
Here, similar to the single-particle quasi-momentum, we can identify $K$ as the c.m. quasi-momentum, which is referred to as the c.m. momentum for brevity.
Although ${K}$ corresponds to the total momentum of all particles, we shall use the terminology ``c.m. momentum" to stress that it is the reciprocal lattice vector with respect to the center-of-mass position.
It is easy to verify that Eq.~\eqref{eqn:Translational_Operator_Eigenstates} obeys
\begin{eqnarray}
\label{eqn:cotranslation_OP_eigenvalues}
 \hat{T}|{K},\beta \rangle &=&\frac{1}{\sqrt{C_\beta}}\sum\limits_{{R}}{{{e}^{i{K}\cdot {R}}}\hat{T}|{R},\beta \rangle }   \nonumber \\
&=& \frac{1}{\sqrt{C_\beta}}\sum\limits_{{R}}{{{e}^{i{K} {R}}}|{R}+1,\beta \rangle }  \nonumber \\
&=& {{e}^{-i{K}}}\frac{1}{\sqrt{C_\beta}}\sum\limits_{{R}}{{{e}^{i{K} {R}}}|{R},\beta \rangle } \nonumber \\
&=& {{e}^{-i{K}}}|{K},\beta \rangle  .
\end{eqnarray}
Consequently, the Hamiltonian can be block-diagonalized into the direct sum of c.m. Bloch Hamiltonians $h\left( K \right)$,
\begin{equation}
\label{eqn:Ham_BlochHam}
\hat{H} = \sum\limits_{{K}}{\sum\limits_{{\beta }',\beta }{|{K},{\beta }'\rangle {{\left[ h({K}) \right]}_{{\beta }',\beta }}\langle {K},\beta |}} = { \bigoplus _K}h\left( K \right),
\end{equation}
where ${{\left[ h({K}) \right]}_{{\beta }',\beta }} =\langle {K},\beta'| \hat{H}|{K},\beta\rangle$.
Thus, the eigenstate can be expressed as the linear combination of c.m. momentum basis $|{K}, \beta\rangle$,
\begin{equation}
\label{eqn:CM_psi_K_u_K_relation}
|\psi _{{K}}^{n}\rangle =\sum\limits_{\beta }{u_{{K},\beta }^{n}|{K},\beta \rangle },
\end{equation}
in which $u^n_{{K} ,\beta }$ is the eigenvector of $h({K})$ satisfying $h\left( {K} \right)|u_{K}^{n}\rangle =E_{{K}}^{n}|u_{{K}}^{n}\rangle  $, and $n$ is the eigenenergy index of $h({K})$.
We will call $|u_{K}^{n}\rangle$ the \emph{c.m. momentum state}.

Under PBC, applying the co-translation operator for $L$ times will yield the same state: $\hat{T}^L|\psi\rangle= |\psi\rangle$.
Hence, from Eq.~\eqref{eqn:cotranslation_OP_eigenvalues}, there is ${K}=2\pi m/{L}, m \in \mathbb{Z}$.
However, for a many-body system of indistinguishable particles, some specific distributions may reduce the needed times of co-translation symmetry to yield the same state.
For example, let us consider two specific states in a simple one-dimensional lattice: (i) the state of all particles distributed uniformly, $|\cdots,1,1,1,\cdots \rangle$, and (ii) the state of the particles distributed uniformly only at odd or even sites, $|\cdots,0,2,0, 2,\cdots \rangle$.
Any co-translation operation will not change this state: $\hat{T}|\cdots,1,1,1,\cdots \rangle = |\cdots,1,1,1,\cdots \rangle$.
According to Eq.~\eqref{eqn:Translational_Operator_Eigenstates}, $|\cdots,1,1,1,\cdots \rangle$ can only be used to construct the c.m. momentum basis with $K=0$.
For the state $|\cdots,0,2,0, 2,\cdots \rangle$, applying the co-translation operator twice will bring the state back to the original state, and therefore it can only be used to construct the c.m. momentum basis with $K=0$ or $K=\pi$.
This fact means that for different c.m. momenta, the number of the eigenstates of the co-translation operator can be different and therefore the matrix dimensions of the Bloch Hamiltonian $h({K})$ may be different.
In dealing with the summation of different relative distributions in Eq.~\eqref{eqn:Ham_BlochHam}, we should carefully distinguish which states should be involved for a certain c.m. momentum.
In addition, the definition of position is essential in constructing the c.m. momentum basis.
The c.m. position can be expressed as $R=R_i+R_{\beta}$ with an integer part $R_i \in \mathbb{Z}$ and a decimal part $R_\beta \in [0, 1)$.
Thus, according to Eq.~\eqref{eqn:Translational_Operator_Eigenstates}, we have
\begin{equation}
\label{eqn:K_2pi_expR}
|{{K}} + 2\pi,\beta \rangle  = {e^{i2\pi   {{R_{\beta}}} }}|{{K}},\beta \rangle.
\end{equation}
In multi-particle systems, $R_\beta$ is generally non-vanishing.
The c.m. Bloch Hamiltonian then satisfies 
\begin{equation}
{\left[ {h(K + 2\pi )} \right]_{\beta ',\beta }} = {\left[ {h(K)} \right]_{\beta ',\beta }}{e^{i2\pi \left( {{R_\beta } - {R_{\beta '}}} \right)}}.
\end{equation}
In matrix notation, there is $h(K + 2\pi ) = {{\cal R}_{2\pi }}h(K){\cal R}_{2\pi }^{ - 1}$, where ${\left[ {{\mathcal{R}_{2\pi}}} \right]_{\beta ',\beta }} = {\delta _{\beta ',\beta }}{e^{ -i 2\pi {R_\beta }}}$.
Hence, the c.m. momentum state satisfies $u_{{{K}} + 2\pi ,\beta }^n = e^{ -i 2\pi {R_\beta }} u_{{{K}} ,\beta }^n$, and we have
\begin{eqnarray}
\label{eqn:u_K_R_K_Relation}
|u_{K + 2\pi }^n\rangle  = {{\cal R}_{2\pi }}|u_K^n\rangle 
\end{eqnarray}
Such a kind of relation is very similar to the discussion for TBC in Sec.~\ref{Sec:sub_TBC}.
On the other hand, it can be checked that the eigenstate always satisfies the periodic condition
\begin{eqnarray}
\label{eqn:psi_K_periodic_2pi}
|\psi _{{{K}} + 2\pi }^n\rangle  &=& \sum\limits_\beta  {u_{{{K}} + 2\pi ,\beta }^n|{{K}} + 2\pi ,\beta \rangle }  \nonumber \\
 &=& \sum\limits_\beta  {u_{{K},\beta}^n|{{K}},\beta \rangle } \nonumber  \\
 &=& |\psi _{{K}}^n\rangle .
\end{eqnarray}
The above discussion for c.m. momentum states is very similar to the band theory for single-particle systems, and can also be generalized to higher-dimensional systems.

\subsection{Connection between the twist angle and the center-of-mass momentum states}
\label{Sec:sub_Connection_TBC_CM}

Below we show how the twist angle connects with the c.m. momentum states.
First, we would like to discuss the general characteristics for both many-body and few-body systems.
It can be noted that the co-translation symmetry is broken under the TBC with boundary gauge, i.e. $[{{{\hat{H}}}_{\text{TBC}}}\left( {\theta } \right), \hat{T}] \ne 0$.
Nevertheless, when the periodic gauge is imposed on the TBC, the system satisfies the PBC, and the co-translation symmetry is restored, i.e. $[{{{\hat{H}}}_{\text{TPG}}}\left( {\theta } \right), \hat{T}] = 0$.
As discussed in Sec.~\ref{Sec:sub_Cotranslation_Sym_and_CM_momentum}, we can block diagonalize the Hamiltonian in this situation and obtain the Bloch Hamiltonian $h({K},{\theta})$ from the c.m. momentum approach discussed above.
Using Eq.~\eqref{eqn:Ham_PBC_theta}, we can find that the matrix elements of the Bloch Hamiltonian satisfy the following relation
\begin{eqnarray}
\label{eqn:Hk_relation0}
&&{{\left[ h\left( {K},{\theta }+2\pi  \right) \right]}_{{\beta }',\beta }}=\langle {K},{\beta }'|{{{\hat{H}}}_{\text{TPG}}}\left( {\theta }+2\pi \right)|{K},\beta \rangle   \nonumber \\
 &&\qquad  = \langle {K},{\beta }'|{{{\hat{U}}}_{2\pi }}{{{\hat{U}}}_{{\theta }}}{{{\hat{H}}}_{\text{TBC}}}\left( {\theta } \right){{{\hat{U}}}_{{\theta }}}^{-1}{{{\hat{U}}}_{2\pi }}^{-1}|{K},\beta \rangle  \nonumber \\
 && \qquad = \langle {K}-N\delta {{K}},{\beta }'|{{{\hat{U}}}_{{\theta }}}{{{\hat{H}}}_{\text{TBC}}}\left( {\theta } \right){{{\hat{U}}}_{{\theta }}}^{-1}|{K}-N\delta{ {K}},\beta \rangle   \nonumber \\
 && \qquad =\langle {K}-N\delta{ {K}},{\beta }'|{{{\hat{H}}}_{\text{TPG}}}\left( {\theta } \right)|{K}-N\delta {{K}},\beta \rangle   \nonumber \\
 && \qquad = {{\left[ h\left( {K}-N\delta {{{K}}},{\theta } \right) \right]}_{{\beta }',\beta }}
\end{eqnarray}
where $\delta {K }= 2\pi /{L}$ is the minimum increment of the c.m. momentum, and we have used the following relation
\begin{eqnarray}
\label{eqn:twist_operator_U2pi_relation}
{{{\hat{U}}}^{-1}_{2\pi }}|{K},\beta \rangle &=&\frac{1}{{\sqrt{C_\beta} }}\sum\limits_R {{e^{iK  R}}{e^{ - i\frac{{2\pi }}{L}\hat x}}|R,\beta \rangle }  \nonumber \\
&=& \frac{1}{\sqrt{C_\beta}}\sum\limits_{{R}}{{{e}^{i{K} {R}}}{{e}^{-i\frac{2\pi }{{{L}}} {NR}}}|{R},\beta \rangle }  \nonumber \\
&=& \frac{1}{\sqrt{C_\beta}}\sum\limits_{{R}}{{{e}^{i\left( {K}-N\delta{ {{K}}} \right) {R}}}|{R},\beta \rangle }\nonumber \\
&=& |{K}-N\delta {{{K}}},\beta \rangle .
\end{eqnarray}
Here $\hat{U}_{2\pi}$ is also called the twist operator \citep{PhysRevLett.79.1110} and it satisfies $\hat T\left( {\hat U_{2\pi }^{ - 1}|K,\beta \rangle } \right) = {e^{i\left( {K - N\delta K} \right)}}\left( {\hat U_{2\pi }^{ - 1}|K,\beta \rangle } \right)$.

From Eq.~\eqref{eqn:Hk_relation0}, we find that the Bloch Hamiltonian $h({K})$ satisfies the following important relation
\begin{equation}
\label{eqn:Hk_relation}
h\left( {K},{\theta }+2\pi  \right)=h\left( {K}-N\delta{ {{K}}},{\theta } \right),
\end{equation}
and the corresponding c.m. momentum state (the eigenstate of $h({K},{\theta})$) will satisfy the following relation:
\begin{equation}
\label{eqn:u_K_Gauge_freedom}
|u_{{{K}} }^n({{\theta }} + 2\pi {})\rangle  = |u_{{K - N{\delta }{{{K}}}}}^n({{\theta }})\rangle.
\end{equation}
This means that the twist angle continuously connects c.m. momentum states in different sectors under the periodic gauge.
Moreover, eigenstates of the Hamiltonian under TBC with periodic gauge \eqref{eqn:Ham_PBC_theta} satisfy a rather different relation
\begin{eqnarray}
\label{eqn:psi_K_PeriodicGauge_relation}
&& |\psi _{{{K}} + N{{\delta }}{{{K}}}}^n\left( {{{\theta }} + 2\pi} \right)\rangle  \nonumber \\
&& \qquad = \sum\limits_\beta  {u_{{{K}} + N{\delta}{{{K}}},\beta }^n\left( {{{\theta }} + 2\pi {}} \right)|{{K}} + N{\delta}{{{K}}},\beta \rangle }   \nonumber \\
&&\qquad   =\sum\limits_\beta  {u_{{{K}},\beta }^n\left( {{\theta }} \right)|{{K}} + N{{\delta }}{{{K}}},\beta \rangle }  \nonumber \\
&& \qquad  = \sum\limits_\beta  {u_{{{K}},\beta }^n\left( {{\theta }} \right){{\hat U}_{2\pi}}|{{K}},\beta \rangle }   \nonumber \\
&& \qquad  = {{\hat U}_{2\pi }}|\psi _{{K}}^n\left( {{\theta }} \right)\rangle  .
\end{eqnarray}

By multiplying ${{\hat U}^{-1}_{{\theta}+2\pi {}}}$ on both sides of Eq.~\eqref{eqn:psi_K_PeriodicGauge_relation}, one can transform the periodic gauge back to the boundary gauge
\begin{equation}
\label{eqn:TBC_eigenstate_flow}
|\tilde \psi _{{{K}} + N{{\delta }}{{{K}}}}^n\left( {{{\theta }} + 2\pi {}} \right)\rangle  = |\tilde \psi _{{K}}^n\left( {{\theta }} \right)\rangle ,
\end{equation}
where $|\tilde \psi _{{K}}^n\left( {{\theta }} \right)\rangle $ is the eigenstates of the TBC Hamiltonian under boundary gauge $ {{{\hat{H}}}_{\text{TBC}}}\left( {\theta } \right)$.
Note that the co-translation symmetry is broken under the boundary gauge, and the c.m. momentum $ K$ is not a good quantum number for $ {{{\hat{H}}}_{\text{TBC}}}\left( {\theta } \right)$.
However, there is still a one-to-one correspondence between the eigenstates of the periodic-gauge Hamiltonian and the boundary-gauge Hamiltonian since they are related by the unitary transformation: $|{\psi _\mu }\left( {{\theta }} \right)\rangle  = {\hat U_{{\theta }}}|{{\tilde \psi }_\mu }\left( {{\theta }} \right)\rangle $.
Here, $\mu$ denotes the index of eigenstates.
Thus, we can still assign the quantum numbers $\{ K, n\}$ to the eigenstates of $ {{{\hat{H}}}_{\text{TBC}}}\left( {\theta } \right)$ such that
\begin{equation}
\label{eqn:K_n_attach_TBC}
|{{\tilde \psi }_\mu }\left( {{\theta }} \right)\rangle  \equiv |\tilde \psi _{{K}}^n\left( {{\theta }} \right)\rangle .
\end{equation}
Under the boundary gauge, a notable consequence of Eq.~\eqref{eqn:TBC_eigenstate_flow} is that when the twist angle $ \theta$ flows from $0$ to $2\pi$, each of the eigenstates will flow \emph{adiabatically} to another eigenstate if $N/L$ is not an integer, although the TBC Hamiltonian under the boundary gauge flows back to the same Hamiltonian.

According to Eq.~\eqref{eqn:Hk_relation}, the eigenenergy will also follow the relation
\begin{equation}
\label{eqn:energy_relation}
E_{{K}}^{n}\left( {\theta }+2\pi  \right)=E_{{K}-N{\delta {{K}}}}^{n}\left( {\theta } \right).
\end{equation}
From Eq.~\eqref{eqn:energy_relation}, we can see that the eigenenergies change continuously from $E_{{K}}^{n}(0)$ to $E_{{K}-N{\delta {{K}}}}^{n}(0)$ when the twist angle ${\theta}$ changes adiabatically $2\pi$.
It has been proven that the finite excitation gap is not affected by the twist angle ${\theta}$ in the thermodynamic limit \citep{PhysRevB.98.155137}.
Physically, it can be understood that the change of the twist angle at the boundary will not affect the bulk when the system is away from the critical point.
This means that if $E_{{K}}^{n}(0)$ is the eigenenergy of the gapped ground state, then the eigenstate whose eigenenergy is $E_{{K}-N\delta{ {{K}}}}^{n}(0)$ also belongs to the ground-state manifold.
Thus, the degeneracy of the gapped ground-state manifold depends on the filling factor $\nu=N/L$.
For example, consider the case of a filling factor $\nu=N/L=p/q$ with $p$ and $q$ being co-prime numbers.
The degeneracy of the ground states must be the multiple of $q$.

The above discussion on the relation between the twist angle and the c.m. momentum is general, and the results can be applied to both many-body and few-body systems.
Actually, this result is in agreement with the celebrated Lieb-Shultz-Mattis (LSM) theorem \citep{LIEB1961407,affleck1986proof,PhysRevLett.78.1984, PhysRevLett.79.1110,PhysRevLett.84.1535, PhysRevB.104.075146}.

\section{Berry phase and Chern number}
\label{Sec:Zak_phase}

In this section, we study the Berry phases and the Chern numbers defined with the TBC and the c.m. momentum states, respectively.
In non-interacting lattice systems, it is known that the Berry phase defined through the single-particle quasi-momentum is related to polarization \citep{PhysRevB.47.1651, PhysRevB.48.4442, RevModPhys.66.899, PhysRevB.87.235123}.
By choosing an appropriate gauge for the Berry connection (Berry vector potential), the Chern number for 2D systems can be expressed as the winding of Berry phase.
The adiabatic change of Berry phase reflects the flow of the current induced by modulation.
The periodic modulation may result in nontrivial Chern number, corresponding to the number of particles being pumped.
Periodic modulations can be a time-dependent lattice potential applied to 1D system, or magnetic flux inserted in the 2D system in a cylinder geometry.
The many-body Berry phase has been studied extensively \citep{PhysRevLett.121.147202}.
In particular, when the system has some symmetries, the Berry phase is used as an order parameter to characterize the symmetry-protected topological phase~\citep{PhysRevB.77.094431, zaletel2014flux}.
Therefore, it is essential to investigate the Berry phase for interacting multi-particle systems.

\subsection{Twisted boundary condition approach to Berry phase}

In this subsection, we will present the Berry phase defined through the twist angle under two different gauges: the boundary gauge and the periodic gauge.
We also discuss the gauge-invariant condition.
Although the two cases are unitarily equivalent, their Berry phases differ by a classical polarization.
Particularly, with the periodic gauge, one can expand the eigenstate up to the first order of $\theta/L$, which allows us to relate the Berry phases respectively defined with the twist angle and the c.m. momentum later.

\subsubsection{Boundary gauge}

Firstly, let us consider a 1D system under TBC with the boundary gauge.
Given a set of target states ${\tilde{\cal G}}\left( \theta  \right) = \left\{ {|{{\tilde \psi }_\mu }\left( \theta  \right)\rangle } \right\}$ (which are gapped to other states), we write them as a vector ${\tilde {\boldsymbol{\Psi }}_\theta } = \left( {|{{\tilde \psi }_1}\left( \theta  \right)\rangle , \cdots ,|{{\tilde \psi }_{\mathcal{N}}}\left( \theta  \right)\rangle } \right)$, in which $\mathcal{N}$ is the number of the target states.
One can use the non-Abelian form to define the Berry phase with the twist angle
\begin{eqnarray}
\label{eqn:ZakPhase_TBC_Definition}
{\phi _{{\mathrm{TBC}}}} &=&  - i \int_0^{2\pi } {\mathrm{d}\theta \;{\mathrm{Tr}}[{\tilde {\mathcal A}}\left( \theta  \right)]}, \nonumber \\
{{\tilde {\mathcal A}}\left( \theta  \right)} &=& {\tilde {\boldsymbol{\Psi }}_\theta }^\dag {\partial _\theta }{\tilde {\boldsymbol{\Psi }}_\theta },
\end{eqnarray}
where the minus sign is imposed for convenience.
Next, it is of importance to discuss when the Berry phase~\eqref{eqn:ZakPhase_TBC_Definition} is gauge-invariant.
Supposing a $U(\mathcal{N})$ gauge transformation, ${\tilde {\boldsymbol{\Psi }}_\theta } \to {\tilde {\boldsymbol{\Psi }}'_\theta } = {\tilde {\boldsymbol{\Psi }}_\theta }{\tilde{\mathcal U}_\theta }$ with $\tilde{\mathcal U}_\theta$ being a continuous function of $\theta$, there is
\begin{equation}
\label{eqn:boundary_gauge_UN_gauge_term}
{\phi _{{\rm{TBC}}}} \to {\phi _{{\rm{TBC}}}} - i\int_0^{2\pi } {{\rm{d}}\theta \;{\rm{Tr}}\left( {{\tilde{\mathcal{ U}}_\theta }^\dag {\partial _\theta }{\tilde{\mathcal{ U}}_\theta }} \right)}.
\end{equation}
Note that the $2\pi$-periodicity of the Hamiltonian ${\hat{H}_{\text{TBC}}}\left( {\theta } \right)$ does not mean its eigenstate will flow back to the original state when the twist angle varies $2\pi$.
According to Eq.~\eqref{eqn:TBC_eigenstate_flow}, when the twist angle varies $2\pi$, the boundary-gauge eigenstates will flow to a different eigenstate if $N/L$ is not an integer.
Therefore, the extra gauge term in Eq.~\eqref{eqn:boundary_gauge_UN_gauge_term} may not be zero.
In other words, it seems that the TBC Berry phase~\eqref{eqn:ZakPhase_TBC_Definition} is gauge-dependent.
As discussed in the previous section, the twist angle will not change the spectral gap in the thermodynamic limit, we have $\tilde{\mathcal G}\left( \theta  + 2\pi \right)=\tilde{\mathcal G}\left( \theta  \right)$.
This means that, under the gapped condition, any target state $|{{\tilde \psi }_{\mu}}{\left( \theta  \right)}\rangle  \in \tilde {\mathcal G}\left( \theta  \right)$ will finally evolve into another eigenstate which still belongs to the same set of target states when the twist angle changes $2\pi$, that is,
\begin{equation}
|{{\tilde \psi }_{\mu}}{\left( \theta +2\pi  \right)}\rangle = |{{\tilde \psi }_{\mu '}}{\left( \theta  \right)}\rangle  \in \tilde {\mathcal G}\left( \theta  \right).
\end{equation}
Hence, we would like to impose ${\tilde {\bf{\Psi }}_{\theta + 2\pi} } = {\tilde {\bf{\Psi }}_\theta } $ in practical calculations, and this leads to ${\tilde{\mathcal{ U}}_{\theta  + 2\pi }} ={\tilde{\mathcal{U}}_\theta }$.
Then, the extra term satisfies
\begin{equation}
\label{eqn:UN_Gauge_integral_winding_number}
\int_0^{2\pi } {{\rm{d}}\theta \;{\rm{Tr}}\left( {\tilde{{\mathcal{U}}_\theta }^\dag {\partial _\theta }{\tilde{\mathcal{U}}_\theta }} \right)}  = 2m\pi,\;\; m\in\mathbb{Z},
\end{equation}
where we have used the fact that this integral produces the winding number of the unitary matrix ${\tilde{\mathcal{U}}_\theta }$.
Therefore, we arrive at the conclusion that the TBC Berry phase modulo $2\pi$ is $U(\mathcal{N})$ gauge-invariant as long as the target states are gapped.

\subsubsection{Periodic gauge}

On the other hand, one may wonder if we can define the Berry phase under the periodic gauge via the same form
\begin{eqnarray}
\label{eqn:ZakPhase_TPBC_Definition}
{\phi _{{\mathrm{TPG}}}} &=&  - i\int_0^{2\pi } {d\theta \;{\mathrm{Tr}}[{ {\mathcal A}}\left( \theta  \right)]}, \nonumber \\
{{ {\mathcal A}}\left( \theta  \right)} &=& { {\boldsymbol{\Psi }}^\dag_\theta } {\partial _\theta }{ {\boldsymbol{\Psi }}_\theta },
\end{eqnarray}
where ${ {\boldsymbol{\Psi }}_\theta } = \left( {|{{ \psi }_1}\left( \theta  \right)\rangle , \cdots ,|{{ \psi }_{\mathcal{N}}}\left( \theta  \right)\rangle } \right)$ corresponds to a set of eigenstates under periodic gauge.
In fact, the Berry phase in Eq.~\eqref{eqn:ZakPhase_TPBC_Definition} is generally not gauge-invariant, since the period of the Hamiltonian under periodic gauge $\hat H_{\mathrm{TPG}}(\theta)$ is not $2\pi$.
Similarly, consider a $U(\mathcal{N})$ gauge transformation ${ {\boldsymbol{\Psi }}_\theta } \to { {\boldsymbol{\Psi }}'_\theta } = { {\boldsymbol{\Psi }}_\theta }{{\mathcal U}_\theta }$ for a set of target states ${{\mathcal G}\left( \theta  \right)} = \left\{ {|{{ \psi }_\mu {\left( \theta  \right)}}\rangle } \right\}$ gapped to other states.
This gauge transformation leads to
\begin{equation}
\label{eqn:periodic_gauge_UN_gauge_term}
{\phi _{{\rm{TPG}}}} \to {\phi _{{\rm{TPG}}}} - i\int_0^{2\pi } {{\rm{d}}\theta \;{\rm{Tr}}\left( {{{\mathcal{ U}}^\dag_\theta } {\partial _\theta }{{\mathcal{ U}}_\theta }} \right)}.
\end{equation}
Apparently, the period of $\mathcal{ U}_\theta$ is not $2\pi$, which means the integral in Eq.~\eqref{eqn:periodic_gauge_UN_gauge_term} modulo $2\pi$ is not necessarily zero, implying that Eq.~\eqref{eqn:ZakPhase_TPBC_Definition} is not gauge invariant.
To make the Berry phase~\eqref{eqn:ZakPhase_TPBC_Definition} gauge-invariant, one can manually fix the gauge.
As discussed in previous sections, the gapped target states satisfy $|{\psi _\mu }\left( {\theta  + 2\pi } \right)\rangle  = {{\hat U}_{2\pi }}|{\psi _{\mu '}}\left( \theta  \right)\rangle $, in which $|{\psi _{\mu '}}\left( \theta  \right)\rangle \in {{\mathcal G}\left( \theta  \right)}$.
With this, we can also impose the following relation
\begin{equation}
{{\boldsymbol{\Psi }}_{\theta  + 2\pi }} = {{\hat U}_{2\pi }}{{\boldsymbol{\Psi }}_\theta },
\end{equation}
and then the $U(\mathcal{N})$ gauge transformation will satisfy the periodic relation ${{\mathcal{U}}_{\theta  + 2\pi }} = {{\boldsymbol{\Psi }}^\dag_{\theta  + 2\pi } }{{\hat U}_{2\pi }}{{\boldsymbol{\Psi }}_\theta }{{\mathcal{ U}}_\theta } = {{\mathcal{ U}}_\theta }$.
Hence, similar to Eq.~\eqref{eqn:UN_Gauge_integral_winding_number}, the extra gauge term will only produce an integer multiple of $2\pi$
\begin{equation}
i\int_0^{2\pi } {{\rm{d}}\theta \;{\rm{Tr}}\left( {{{\mathcal{ U}}^\dag_\theta } {\partial _\theta }{{\mathcal{ U}}_\theta }} \right)} = 2m\pi,\;\; m\in \mathbb{Z},
\end{equation}
and the Berry phase under periodic gauge [Eq.~\eqref{eqn:ZakPhase_TPBC_Definition}] is gauge-invariant modulo $2\pi$ now.
This is particularly useful in practical computation.

\subsubsection{Relation of the Berry phases for different gauges}
\label{Sec:subsub_Relation_TBC_different_Gauges}
By using the relation between the eigenstates under the two different gauges, ${\boldsymbol{\Psi} _\theta } = {\hat U_\theta }{\tilde{\boldsymbol{\Psi}} _\theta } $, the Berry phase \eqref{eqn:ZakPhase_TPBC_Definition} becomes
\begin{eqnarray}
\label{eqn:Zak_phase_PBC_to_TBC}
{\phi _{{\mathrm{TPG}}}} &=&  - i\int_0^{2\pi } {\mathrm{d}\theta \;{\mathrm{Tr}} [{{ {\mathcal A}}}\left( \theta  \right)]}  \nonumber \\
&=& - i\int_0^{2\pi } {\mathrm{d}\theta \;{\mathrm{Tr}} [{{\tilde {\mathcal A}}}\left( \theta  \right)]}  \nonumber \\
&& - i\int_0^{2\pi } {{\rm{d}}\theta \;{\rm{Tr}}\left[ {\tilde {\boldsymbol{\Psi }}_\theta ^\dag \hat U_\theta ^\dag \left( {{\partial _\theta }{{\hat U}_\theta }} \right){{\tilde {\boldsymbol{\Psi }}}_\theta }} \right]}   \nonumber \\
&=&  {\phi _{{\mathrm{TBC}}}}+2\pi \bar{P},\nonumber \\
\end{eqnarray}
where 
\begin{equation}
\label{eqn:classical_polarization}
\bar P = \frac{1}{{2\pi L}}\int_0^{2\pi } {{\rm{d}}\theta \;{\rm{Tr}}\left( {\tilde {\boldsymbol{\Psi }}_\theta ^\dag \hat x{{\tilde {\boldsymbol{\Psi }}}_\theta }} \right)} 
\end{equation}
corresponds to the classical polarization averaged over the twist angle.
The formula~\eqref{eqn:Zak_phase_PBC_to_TBC} reveals that the two TBC Berry phases differ by a classical polarization, which is consistent with the results of Ref.~\citep{PhysRevX.8.021065} where the target state only consists of one unique eigenstate.
On the other hand, it is known that the definition of the position is somewhat arbitrary due to the TBC.
Eq.~\eqref{eqn:Zak_phase_PBC_to_TBC} suggests that the TBC Berry phases under either the boundary gauge or the periodic gauge are affected by the choice of the position operator $\hat x$.
Apparently, the TBC Berry phase under the boundary gauge [Eq.~\eqref{eqn:ZakPhase_TBC_Definition}] does not involve any position information except for the determination of boundary.
It should be irrelevant to how the position operator is defined.
Thus, we can conclude that only the TBC Berry phase under the periodic gauge ($\phi_{\mathrm{TPG}}$) depends on the definition of the position operator.
This is similar to the single-particle situation, where the Berry phase can be split into the inter-cellular and intra-cellular parts \citep{PhysRevB.95.035421}.
Correspondingly, the inter-cellular Berry phase in the single-particle case corresponds to the TBC Berry phase~\eqref{eqn:ZakPhase_TBC_Definition}, and the intra-cellular Berry phase corresponds to the classical polarization part.
Unlike Ref.~\citep{PhysRevB.95.035421}, the relation obtained here is purely based on the TBC, and can be applied to both single-particle and multi-particle systems without requiring translation symmetry.

\subsubsection{Perturbative analysis for periodic gauge}
\label{Secsubsection_perturbative_analysis}

To see the perturbative nature of the TBC Berry phase, let us consider a system described by the Hamiltonian under TBC with periodic gauge, as introduced in Eq.~\eqref{eqn:Ham_PBC_theta}.
In this condition, the co-translation symmetry is preserved.
We can label the eigenstate by good quantum numbers: $|{\psi _\mu }\left( \theta  \right)\rangle  \equiv |\psi _K^n\rangle $.
Since the tunneling is assumed to be finite-range, it can be seen from the Hamiltonain \eqref{eqn:Ham_PBC_theta} that the twist angle always appears as a extremely small quantity $\theta/L$ in the thermodynamic limit.
Hence, provided the tunneling is finite-range, the eigenstate can be expanded in terms of ${\theta}/{L}$:
 \begin{equation}
 \label{eqn:PBC_psi_expand}
 |\psi _K^n\left( \theta  \right)\rangle  = |\psi _K^n\left( 0 \right)\rangle  + \frac{\theta }{L}|{\partial _{\theta /L}}\psi _K^n\left( \theta  \right)\rangle_{\theta=0}  + O\left( {\frac{1}{L^2}} \right).
 \end{equation}
 %
%
By taking derivatives for both sides, we have
\begin{eqnarray}
{\partial _\theta }|\psi _K^n\left( \theta  \right)\rangle  &=& \frac{1}{L}|{\partial _{\theta /L}}\psi _K^n\left( \theta  \right){\rangle _{\theta  = 0}} + O\left( {\frac{1}{{{L^2}}}} \right),  \nonumber \\
 &=&|{\partial _\theta }\psi _K^n\left( \theta  \right){\rangle _{\theta  = 0}} + O\left( {\frac{1}{{{L^2}}}} \right).
\end{eqnarray}
Therefore, up to the first order of $\frac{\theta}{L}$, we obtain
\begin{eqnarray}
 \label{eqn:PBC_BerryConnection_expand}
&&\langle \psi _{K'}^{n'}\left( \theta  \right)|{\partial _\theta }\psi _K^n\left( \theta  \right)\rangle   \nonumber \\
&& \qquad = {\langle \psi _{K'}^{n'}\left( 0 \right)|{\partial _{\theta }}\psi _K^n\left( \theta  \right)\rangle _{\theta  = 0}} + O\left( {\frac{1}{L^2}} \right).
\end{eqnarray}
This means that, up to the first order of $\frac{1}{L}$, the quantity $\langle \psi _{K'}^{n'}\left( \theta  \right)|{\partial _\theta }\psi _K^n\left( \theta  \right)\rangle$ is independent of the twist angle $\theta$.
Then, we can set $\theta=2\pi$ in Eq.~\eqref{eqn:PBC_psi_expand}
\begin{equation}
|{\partial _\theta }\psi _K^n\left( \theta  \right){\rangle _{\theta  = 0}} = \frac{1}{{2\pi }}\left[ {|\psi _K^n\left( {2\pi } \right)\rangle  - |\psi _K^n\left( 0 \right)\rangle } \right] + O\left( {\frac{1}{L^2}} \right),
\end{equation}
and therefore
\begin{eqnarray}
\langle \psi _{K'}^{n'}\left( \theta  \right)|{\partial _\theta }\psi _K^n\left( \theta  \right)\rangle  &&= \frac{1}{{2\pi }}\langle \psi _{K'}^{n'}\left( 0 \right)|\psi _K^n\left( {2\pi } \right)\rangle \nonumber \\
&& - \frac{1}{{2\pi }}{\delta _{n',n}}{\delta _{K',K}} + O\left( {\frac{1}{L^2}} \right).
\end{eqnarray}
Furthermore, according to Eq.~\eqref{eqn:psi_K_PeriodicGauge_relation}, one can use $|\psi _K^n\left( {2\pi } \right)\rangle  = {\hat{U}_{2\pi }}|\psi _{K - N\delta K}^n\left( 0 \right)\rangle $ and find
\begin{eqnarray}
\langle \psi _{K'}^{n'}\left( \theta  \right)|{\partial _\theta }\psi _K^n\left( \theta  \right)\rangle  &&= \frac{1}{{2\pi }}\langle \psi _{K'}^{n'}\left( 0 \right)|{\hat{U}_{2\pi }}|\psi _{K - N\delta K}^n\left( 0 \right)\rangle \nonumber \\
&& - \frac{1}{{2\pi }}{\delta _{n',n}}{\delta _{K',K}} + O\left( {\frac{1}{L^2}} \right).
\end{eqnarray}
Hence, we can approximate the Berry connection as
\begin{equation}
\label{eqn:Connection_A_approximation_M}
{\mathcal A}\left( \theta  \right) \approx {\mathcal M} - I_{\mathcal{N}}
\end{equation}
where $I_\mathcal{N}$ is a $\mathcal{N} \times \mathcal{N}$ identity matrix and $\mathcal M$ is a $\mathcal{N} \times \mathcal{N}$ matrix with elements
\begin{equation}
\label{eqn:matrix_M_element}
{\mathcal{M}_{(n',K'),(n,K)}} =\langle \psi _{K'}^{n'}(0)|{\hat{U}_{2\pi }}|\psi _{K - N\delta K}^n(0)\rangle. 
\end{equation} 
In the above, $\mathcal{N}$ is the number of target states and we have dropped the notation of the twist angle since $\theta=0$.
This means that the Berry connection $\mathcal{A}(\theta)$ in Eq.~\eqref{eqn:ZakPhase_TPBC_Definition} is independent of the twist angle $\theta$ up to the first order of $\frac{1}{L}$.
Similarly, up to the first order of $\frac{1}{L}$, we can approximate the classical polarization \eqref{eqn:classical_polarization} as
\begin{equation}
\label{eqn:classical_polarization_approximation}
\bar P \approx  \sum\limits_{n,K} {\langle \psi _K^n|\frac{{\hat x}}{L}|\psi _K^n\rangle }.
\end{equation}

We also find that, in the thermodynamic limit, the matrix $\mathcal{M}$ is approximately a unitary matrix in the subspace spanned by target states, see the detailed discussion in Appendix.~\ref{appendix:unitarity_of_M}.
Then, in the thermodynamic limit, the TBC Berry phase~\eqref{eqn:ZakPhase_TPBC_Definition} can be written as \citep{higham2008functions},
\begin{equation}
\label{eqn:M_matrix_polarization}
{\phi _{{\mathrm{TPG}}}} =  {\mathrm{Im}}\left[ \mathrm{Tr} ({\mathcal{ M} - I_{\mathcal{N}}}) \right] \approx \mathrm{Arg}\left[\det \left( \mathcal{M}\right)\right].
\end{equation}
A similar approximation has been used in Ref.~\citep{PhysRevB.103.224208}.
The above formula~\eqref{eqn:M_matrix_polarization} is related to the polarization formula proposed by Resta~\citep{PhysRevLett.80.1800}, which has been widely applied to investigate the polarization of various systems, from non-interacting \citep{RevModPhys.84.1419, PhysRevLett.113.046802,Science362,PhysRevB.96.245115} to interacting \citep{PhysRevB.103.035112, PhysRevLett.126.050501} systems.
It can be found that
\begin{equation}
\label{eqn:M_matrix_representation}
{\mathcal M} = {\boldsymbol{\Psi }}_0^\dag {{\hat U}_{2\pi }}{{\boldsymbol{\Psi }}_0}{\mathcal S},
\end{equation}
where ${\mathcal S}$ is an orthogonal matrix that permutes the order of the eigenstates in ${{\boldsymbol{\Psi }}_0}$ depending on the flow of the target states. 
Since the orthogonal matrix satisfies $\det ({\mathcal S})=\pm 1$, the Berry phases obtained from the TBC method and the Resta's formula may at most have a $\pi$ phase difference.

\subsection{Center-of-mass momentum approach to Berry phase}

Next, let us discuss the Berry phase defined through c.m. momentum states.
In few-body systems, the filling number tends to zero $\nu=N/L \to 0$ while the total particle number $N$ is fixed. 
In analogy to the single-particle system, the band structure appears, as demonstrated in Fig.~\ref{fig:FIG_Spectrum_illustration} (b).
Hence, in the same fashion, it is desirable to define the Berry phase through c.m. momentum states
\begin{equation}
\label{eqn:Berry_Zak_CM_integral}
\phi =  i\int_0^{2\pi } {{\rm{d}}K\;{\rm{Tr}}\left( {{A_K}} \right)} ,
\end{equation}
in which ${\left[ {{A_K}} \right]_{m,n}} = \langle u_K^m|{\partial _K}u_K^n\rangle $.
To guarantee the gauge invariance, we have to impose $ |u_{2\pi}^n\rangle =\mathcal{R}_{2\pi}|u_0^n\rangle $ according to Eq.~\eqref{eqn:u_K_R_K_Relation}.
Eq.~\eqref{eqn:Berry_Zak_CM_integral} reflects the geometric phase gained by the few-body system after traveling through the Brillouin zone adiabatically.
This reveals the topological property of the Brillouin manifold with respect to the c.m. momentum state.
In particular, Eq.~\eqref{eqn:Berry_Zak_CM_integral} has been successfully used to investigate the topological properties of few-body bound states  \citep{PhysRevB.96.195134, Qin_2018, PhysRevA.95.063630,PhysRevA.101.023620}.
However, Eq.~\eqref{eqn:Berry_Zak_CM_integral} cannot be applied to the many-body system, as the number of gapped ground states is finite, and therefore we cannot use the integral formulation.
To unify the c.m. approach for  few-body and many-body systems, it is desirable to use the Wilson loop to calculate the Berry phase.
Recall that in the single-particle case, the Wilson loop reads as $ {{{\mathcal{W}}^{(1)}_{K \to K- 2\pi }}} =  {{F^{(1)}_K}} {{F^{(1)}_{K-\delta K}}}\cdots  {{F^{(1)}_{K-2\pi+\delta K}}}$, in which ${\left[ {{F^{(1)}_K}} \right]_{n',n}} = \langle u_K^{n'}|u_{K - \delta K}^n \rangle$ and the superscript denotes the particle number.
For the $N$-particle system with the filling number $\nu =N/L=p/q$, we propose that the \emph{$N$-particle Wilson loop} should be modified as 
\begin{equation}
 {{{\mathcal{W}}^{(N)}_{K \to K- 2p\pi }}} = {{F^{(N)}_K}}{{F^{(N)}_{K-N\delta K}}} \cdots {{F^{(N)}_{K-2p\pi + N\delta K}}},
\end{equation}
in which ${\left[ {{F^{(N)}_K}} \right]_{n',n}} = \langle u_K^{n'}|u_{K - N \delta K}^n \rangle$.
We can consider the $N$-particle Wilson loop as a generalization of the single-particle Wilson loop.
The increment of the quasi-momentum is $N\delta K$ for the $N$-particle system, and the range of the Wilson loop depends on the filling factor $\nu =N/L=p/q$.
The Brillouin zone is now expanded $p$ times to complete the loop.
The c.m. Berry phase is therefore defined as
\begin{equation}
\label{eqn:Berry_phase_CM_WilsonLoop}
\phi_{\rm{c.m.}} (K)= {{\rm{Arg}}\left[ {\det \left( {{{\mathcal{W}}^{(N)}_{K \to K - 2p\pi }}} \right)} \right]},
\end{equation}
where $K$ is the starting point of the Wilson loop.
Note that one should impose the relation in Eq.~\eqref{eqn:u_K_R_K_Relation} to guarantee the gauge invariance.

\subsection{Connection between the TBC Berry phase and the c.m. Berry phase}
\label{sec:sub_connection_between_TBC_and_CM}

\begin{figure}
  \includegraphics[width = \columnwidth ]{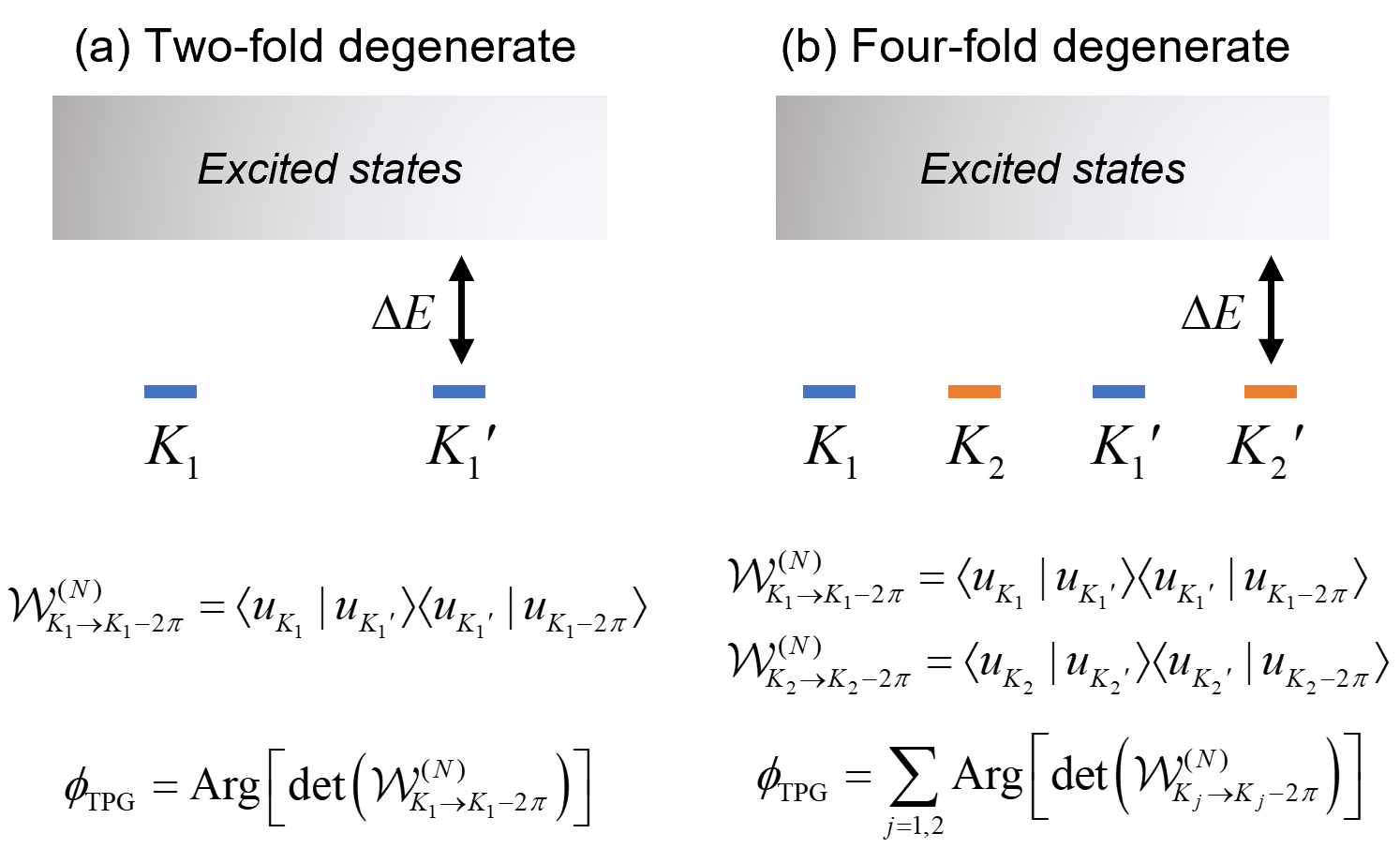}
  \caption{\label{fig:FIG_illustration_4fold_GS}
 Illustrative example for the TBC Berry phase and the multi-particle Wilson loop.
 (a) and (b) show two different cases of the target states at $\nu=1/2$ filling.
 The color distinguishes two different sets of the c.m. momentum states, as defined in Eq.~\eqref{eqn:K_subset}.
 In each set, c.m. momenta only differ by integer multiple of $N\delta K$.
 In (a), the target states only form one single multi-particle Wilson loop.
 In (b), the target states form two multi-particle Wilson loops.
  }
\end{figure}

From Eq.~\eqref{eqn:matrix_M_element}, one can find that the matrix $\mathcal{M}$ has a block-diagonal structure
\begin{eqnarray}
&& {\mathcal{M}_{(n',K'),(n,K)}}   \nonumber \\
&& \qquad = \langle \psi _{K'}^{n'}|{\hat U_{2\pi }}|\psi _{K - N\delta K}^n\rangle  \nonumber \\
&& \qquad = \sum\limits_{\beta ',\beta } {{{\left( {u_{K',\beta '}^{n'}} \right)}^*}u_{K,\beta }^n\langle K' ,\beta |{{\hat U}_{2\pi }}|K - N\delta K,\beta \rangle } \nonumber \\
&& \qquad =  \langle u_K^{n'}|u_{K - N\delta K}^n\rangle {\delta _{K',K}},
\end{eqnarray}
we can write $\mathcal{M} = \bigoplus_K F^{(N)}_K$ with ${\left[ {{F^{(N)}_K}} \right]_{n',n}} = \langle u_K^{n'}|u_{K - N\delta K}^n \rangle$ and the indices running over all c.m. momenta of the target states $K\in \{K_{\rm{target}} \}$.
For convenience, we use the superscript $(N)$ in $F^{(N)}_K$ to emphasize that the increment of the c.m. momentum is $N\delta K$.
Then, under the periodic gauge, the TBC Berry phase can be written as
\begin{eqnarray}
\label{eqn:ZakPhase_TBC_WilsonLoop_K}
{\phi _{{\mathrm{TPG}}}} &=& \mathrm{Arg}\left[\det \left( \mathcal{M}\right)\right] \nonumber \\
&=& \sum\limits_{K\in \{K_{\rm{target}}\}} {{\mathrm{Arg}}  \left[{\mathrm{det}}\left({F^{(N)}_K}\right)\right]}.
\end{eqnarray}
As the matrix $\mathcal{M}$ is defined via the states for $\theta=0$ (i.e. the states for the Hamiltonian under PBC), the above formula implies that the TBC Berry phase can be equivalently formulated by the c.m. momentum states under PBC.
As discussed in Sec.~\eqref{Sec:sub_Connection_TBC_CM}, the twist angle continuously connects certain c.m. momentum sectors.
We can collect these c.m. momenta to form a subset
\begin{equation}
\label{eqn:K_subset}
\{ {{\tilde K}_j}\}  = \left\{ {K\left| {K = {K_j} - 2n\pi \frac{p}{q},\;n = 0,1, \cdots ,q - 1} \right.} \right\},
\end{equation}
where $K_j$ is one of the c.m. momenta in the target states.
Hence, the c.m. momenta in the target states can be written as the union of these subsets $\{ {K_{{\rm{target}}}}\}  = { \cup} _j\{ {{\tilde K}_j}\} $.
Based upon this arrangement, Eq.~\eqref{eqn:ZakPhase_TBC_WilsonLoop_K} can be written as
\begin{eqnarray}
\label{eqn:ZakPhase_PBC_WilsonLoop_K}
{\phi _{{\mathrm{TPG}}}}& =& \sum\limits_j\sum\limits_{K \in \{ \tilde{K}_j\}} {{\mathrm{Arg}}  \left[{\mathrm{det}}\left({F^{(N)}_{K}}\right)\right]} \nonumber \\
&=&  \sum\limits_j {{\rm{Arg}}\left[ {\det \left( {{{\mathcal{W}}^{(N)}_{K_j \to K_j - 2p\pi }}} \right)} \right]} \nonumber \\
&=&  \sum\limits_j {\phi_{\rm{c.m.}}(K_j)}.
\end{eqnarray}
In this manner, we have proven that the TBC Berry phase is related to the Berry phase defined through the c.m. momentum state [Eq.~\eqref{eqn:Berry_phase_CM_WilsonLoop}].
To better illustrate the relation in Eq.~\eqref{eqn:ZakPhase_PBC_WilsonLoop_K}, let us consider a fictitious $\nu=1/2$ systems in one dimension.
Two specific cases are assumed: (i) there appear two-fold degenerate ground states [Fig.~\ref{fig:FIG_illustration_4fold_GS} (a)]; (ii) four-fold degenerate ground states [Fig.~\ref{fig:FIG_illustration_4fold_GS} (b)].
For two-fold degenerate ground states, all these c.m. momentum states are connected by the twist angle.
The TBC Berry phase only consists of one single multi-particle Wilson loop.
For four-fold degenerate ground states, the ground states can be classified into two sets, and the c.m. momentum states in each set are connected by the twist angle.
These c.m. momentum states will form two multi-particle Wilson loops, respectively, and the TBC Berry phase is contributed from these two parts according to Eq.~\eqref{eqn:ZakPhase_PBC_WilsonLoop_K}.

\subsection{Discussions for few-body systems and many-body systems}
For a few-body system in the thermodynamic limit, Eq.~\eqref{eqn:Berry_phase_CM_WilsonLoop} can be equivalently written as
\begin{equation}
\phi_{\rm{c.m.}} =  i\int_0^{2p\pi } {{\rm{d}}K\;{\rm{Tr}}\left( {{A_K}} \right)} .
\end{equation}
It can be seen that the $N$-particle Wilson loop in Eq.~\eqref{eqn:Berry_phase_CM_WilsonLoop} covers $p$ Brillouin zones.
Thus, it is $p$ times the result in Eq.~\eqref{eqn:Berry_Zak_CM_integral}, and these two methods are equivalent up to a constant factor.
According to Eq.~\eqref{eqn:u_K_R_K_Relation}, different choices of position definition lead to different relations of unitary transformation for the c.m. momentum state.
Similar to the derivation in Sec.~\ref{Sec:subsub_Relation_TBC_different_Gauges}, it can be proved that the change of position definition only leads to an extra constant term.
On the other hand, it is proved that this c.m. momentum Berry phase is related to the c.m. position of the multi-particle Wannier state \citep{PhysRevA.95.063630, PhysRevA.101.023620}.
This is also very similar to the single-particle case.
In particular, the few-body bound state can be considered as an effective single particle \citep{PhysRevB.96.195134, Qin_2018, PhysRevA.95.063630, PhysRevA.97.013637,PhysRevA.97.013637,PhysRevA.101.023620,Marques_2018,PhysRevResearch.2.013348, PhysRevResearch.2.033267,Mei_2019,Malki_2020}.
The relative distribution is treated as the internal degree of freedom of the effective single particle.
It can also be seen that, for few-body systems, the integral with respect to the twist angle is equivalent to the integral with respect to the c.m. momentum when calculating the Berry phase.
For a many-body system, the $N$-particle Wilson loop only consists of finite eigenstates in the thermodynamic limit, as discussed above.
According to Eq.~\eqref{eqn:u_K_Gauge_freedom}, we can find that the c.m. momentum states are connected by the twist angle.
The perturbative analysis in Sec.~\ref{Secsubsection_perturbative_analysis} implies that, in the many-body case, all these c.m. momentum states only differ by a phase up to the first order of $1/L$ as long as they are connected by the twist angle.
Hence, we do not have to calculate the full Wilson loop in this condition.
It is sufficient to only calculate the overlap between the starting point $K_j$ and the ending point $K_j-2p\pi$:
\begin{equation}
\label{eqn:BerryPhase_WilsonLoop_Manybody_simplification}
{{\mathcal {W}}^{(N)}_{K_j \to K_j - 2p\pi }} \approx  F_{{K_j}}^{\left( {qN} \right)},
\end{equation}
where ${\left[ {F_{{K_j}}^{\left( {qN} \right)}} \right]_{m,n}} = \langle u_{{K_j}}^m|u_{{K_j} - qN\delta K}^n\rangle  = \langle u_{{K_j}}^m|u_{{K_j} - 2p\pi }^n\rangle $ (recall that $\delta K=2\pi / L$ and $N/L = p/q$).
According to Eq.~\eqref{eqn:u_K_R_K_Relation}, there is $|u_{{K_j} - 2p\pi }^n\rangle  =  {{{\mathcal R}_{ - 2p\pi }}} |u_{{K_j}}^n\rangle $.
Hence, we have a rather simple expression
\begin{equation}
    {\left[ {F_{{K_j}}^{\left( {qN} \right)}} \right]_{m,n}} = \langle u_{{K_j}}^m| {{{\mathcal R}_{ - 2p\pi }}}|u_{{K_j} }^n\rangle.
\end{equation}
This result means that we only need to compute one of the target states to obtain the Berry phase, which is more efficient when the target states are multi-fold degenerate.

\subsection{Chern number}
\label{Sec:Chern_number}

Having investigated the Berry phase, below we show that the Chern number can be written as the winding of the Berry phase.
Let us consider a 1D system with a time-periodic modulation, dubbed the (1+1)D system, since the time periodic modulation can be viewed as an artificial dimension~\citep{RevModPhys.88.035005}.
In such a system, the modulation adiabatically changes the lattice potential and results in an adiabatic current~\citep{PhysRevB.27.6083}.
After a pumping period, the Hamiltonian returns to its original form.
Under the TBC, we consider a set of gapped target states $\tilde {\boldsymbol{\Psi }}\left( {\theta ,\tau } \right) = \left( {|{{\tilde \psi }_1}\left( {\theta ,\tau } \right)\rangle , \cdots ,|{{\tilde \psi }_{\cal N}}\left( {\theta ,\tau } \right)\rangle } \right)$.
Thus we have $\tilde {\boldsymbol{\Psi }}\left( {\theta  + 2\pi ,\tau } \right) = \tilde {\boldsymbol{\Psi }}\left( {\theta ,\tau  + T} \right) = \tilde {\boldsymbol{\Psi }}\left( {\theta ,\tau } \right)$.
The Chern number can be written as~\cite{niu1984quantised}
\begin{eqnarray}
\label{eqn:Chern_Num_nonAbelian_TPBC}
C_{\mathrm{(1+1)D}} & =& - \frac{1}{{2\pi }}\int\nolimits_0^{T } {\mathrm{d}{\tau}\int\nolimits_0^{2\pi } {\mathrm{d}{\theta}\;{\mathrm{Tr}}\left[\tilde{\mathcal F}\left( {{\theta , \tau}} \right) \right]}} ; \nonumber\\
\tilde{\mathcal F}\left( { \theta ,\tau} \right) &=& i{\partial _\tau }{\tilde{\mathcal A}_\theta }\left( \theta ,\tau \right) - i{\partial _\theta }{\tilde{\mathcal A}_\tau }\left( \theta ,\tau \right)  \nonumber\\
&&+ \left[ {{\tilde{\mathcal A}_\tau }\left( \theta ,\tau \right),{{\tilde{\mathcal A}}_\theta }\left( \theta ,\tau \right)} \right].
\end{eqnarray}
The commutator term in the non-Abelian Berry curvature $\tilde{\mathcal F}\left( { \theta ,\tau} \right) $ will vanish after the trace operation, therefore the Chern number reads as
\begin{eqnarray}
{C_{({\rm{1 + 1}}){\rm{D}}}} &=&- \frac{1}{{2\pi }}\int_0^T {{\rm{d}}\tau \int_0^{2\pi } {{\rm{d}}\theta \;i{\partial _\tau } \left[ {\rm{Tr}}\left( {{{\tilde {\cal A}}_\theta }} \right) \right]} } \nonumber \\
&&  + \frac{1}{{2\pi }}\int_0^T {{\rm{d}}\tau \int_0^{2\pi } {{\rm{d}}\theta \;i{\partial _\theta }\left[{\rm{Tr}}\left( {{{\tilde {\cal A}}_\tau }} \right)\right]} } ,
\end{eqnarray}
where we have exchanged the orders of trace and partial derivative operations.
Now, if one integrates the twist angle $\theta$ first, the second term will vanish   
\begin{eqnarray}
\int_0^{2\pi } {{\rm{d}}\theta \;i{\partial _\theta }\left[ {{\rm{Tr}}\left( {{{\tilde {\cal A}}_\tau }} \right)} \right]}  &=& i{\rm{Tr}}\left[ {{{\tilde {\cal A}}_\tau }\left( {2\pi ,\tau } \right) - {{\tilde {\cal A}}_\tau }\left( {0,\tau } \right)} \right] \nonumber \\
&=& 0 ,
\end{eqnarray}
since the Berry connection $ {{\tilde {\cal A}}_\tau }\left( {\theta,\tau } \right)$ is a single-valued periodic function.
Finally, by exchanging the orders of integral and partial derivative, one can find that the Chern number can be written as a winding of the Berry phase
\begin{eqnarray}
\label{eqn:ChernNum_1p1D_formula}
{C_{({\rm{1 + 1}}){\rm{D}}}} &=& - \frac{1}{{2\pi }}\int_0^T {{\rm{d}}\tau \int_0^{2\pi } {{\rm{d}}\theta \;i{\partial _\tau }\left[ {{\rm{Tr}}\left( {{{\tilde {\cal A}}_\theta }} \right)} \right]} }  \nonumber \\
& =& \frac{1}{{2\pi }}\int_0^T {{\rm{d}}\tau \;{\partial _\tau }\left\{- {\int_0^{2\pi } {{\rm{d}}\theta \;i\left[ {{\rm{Tr}}\left( {{{\tilde {\cal A}}_\theta }} \right)} \right]} } \right\}}  \nonumber  \\
 &=& \frac{1}{{2\pi }}\int_0^T {{\rm{d}}\tau \;{\partial _\tau }\left[  {{\phi _{{\rm{TBC}}}}\left( \tau  \right)} \right]} ,
\end{eqnarray}
where the minus sign is relevant to the form of the twist angle.
As for 2D systems under TBC, one can consider one of the twist angles as a modulation parameter, which leads to the same result.
Eq.~\eqref{eqn:ChernNum_1p1D_formula} suggests that the Chern number can be derived from the Berry phase.
According to Eq.~\eqref{eqn:Zak_phase_PBC_to_TBC}, we know that the Berry phase under periodic gauge and boundary gauge only differ by a classical polarization $\bar{P}$, which vanishes after a pumping cycle:
\begin{equation}
\int\nolimits_0^T {{\rm{d}}\tau \;{\partial _\tau }\bar P\left( \tau  \right)}  = \bar P\left( T \right) - \bar P\left( 0 \right) = 0.
\end{equation}
Since we have established the relation between the TBC Berry phase and the c.m. Berry phase in Sec.~\ref{sec:sub_connection_between_TBC_and_CM}, we can use the c.m. momentum state to equivalently calculate the Chern number.

\section{Demonstration via interacting Aubry-Andr{\'e}-Harper model}
\label{Sec:Numerical_verification}
In this section, we employ a simple but typical 1D topological model, the Aubry-Andr{\'e}-Harper (AAH) model ~\citep{Harper_1955, aubry1980analyticity}, to demonstrate the above general framework for both many-body and few-body situations.
The AAH model consists of spatial modulations on the either tunneling strength or on-site potentials.
The non-interacting AAH model can be viewed as a reduction of 2D Hofstadter model \citep{PhysRevLett.109.116404, PhysRevLett.110.180403}, and has been realized in various experimental platform \citep{roati2008anderson, PhysRevLett.109.106402,PhysRevLett.120.160404}.
Thouless points out that adiabatic cyclic modulation in the 1D lattice may lead to quantized pumping of particles, provided that the spectral gap is preserved \citep{PhysRevB.27.6083, niu1984quantised}.
The topological origin of this quantized pumping much resembles to the quantum Hall effect.
Later, the charge pumping is associated with the modern theory of polarization \citep{PhysRevB.47.1651,PhysRevB.48.4442, RevModPhys.66.899}.
By changing the modulation phase, one is able to achieve the well-known Thouless pumping \citep{PhysRevLett.111.026802, PhysRevB.27.6083} via the AAH model.
Notably, a special case of AAH model, called the Rice-Mele model \citep{PhysRevLett.49.1455}, has been experimentally realized by loading ultracold atoms into a superlattice~\citep{lohse2016thouless,nakajima2016topological}, in which the quantized topological pumping is observed.
Recently, the interaction effect in such kind of model has been experimentally studied \citep{de2019observation,arxiv.2204.06561}.
During the pumping cycle, the Berry phase will change with the modulation parameter, corresponding to the existence of adiabatic current.
Therefore, it is desirable to calculate and compare the TBC and c.m. Berry phases in the same system.

Below we only consider the spatial modulation on the tunneling strength, dubbed the off-diagonal AAH model.
The Hamiltonian reads as
\begin{eqnarray}
\label{eqn:AAH_model_Ham_nonInt}
{{\hat H}_{\rm{AAH}}}(\Phi ) &=&  - \sum\limits_j {\left( {{t_j}(\Phi )\hat a_{j + 1}^\dag {{\hat a}_j} + \rm{H.c.}} \right)} 
\end{eqnarray}
in which $\hat a_i$ ($\hat a_i^\dag$) are the annihilation (creation) operators of hard-core bosons, $\tau$ is the modulation parameter.
The hopping strengths and the on-site energies are modulated respectively according to ${t_j}\left( \Phi  \right) = t_0 [1- \lambda\cos \left( {2\pi b j  + \Phi } \right)]$, in which $\phi$ is the modulation phase and $b$ controls the period of the tunneling strength.
Moreover, there may appear gapped eigenstates in both many-body and few-body situations if interaction among particles is added, and we can calculate the Berry phases for them.
In the following, the modulated tunneling strength is chosen as $\lambda=0.5 t_0$ with $t_0=1$, which is essential to open the energy gap.
We also set $b=1/3$ so that the system's period is $1/b=3$.
Within these parameters, we can obtain three energy bands with Chern number $C=\{ -1,\ +2, \ -1 \}$ in the single-particle case.

\begin{figure}
  \includegraphics[width = \columnwidth ]{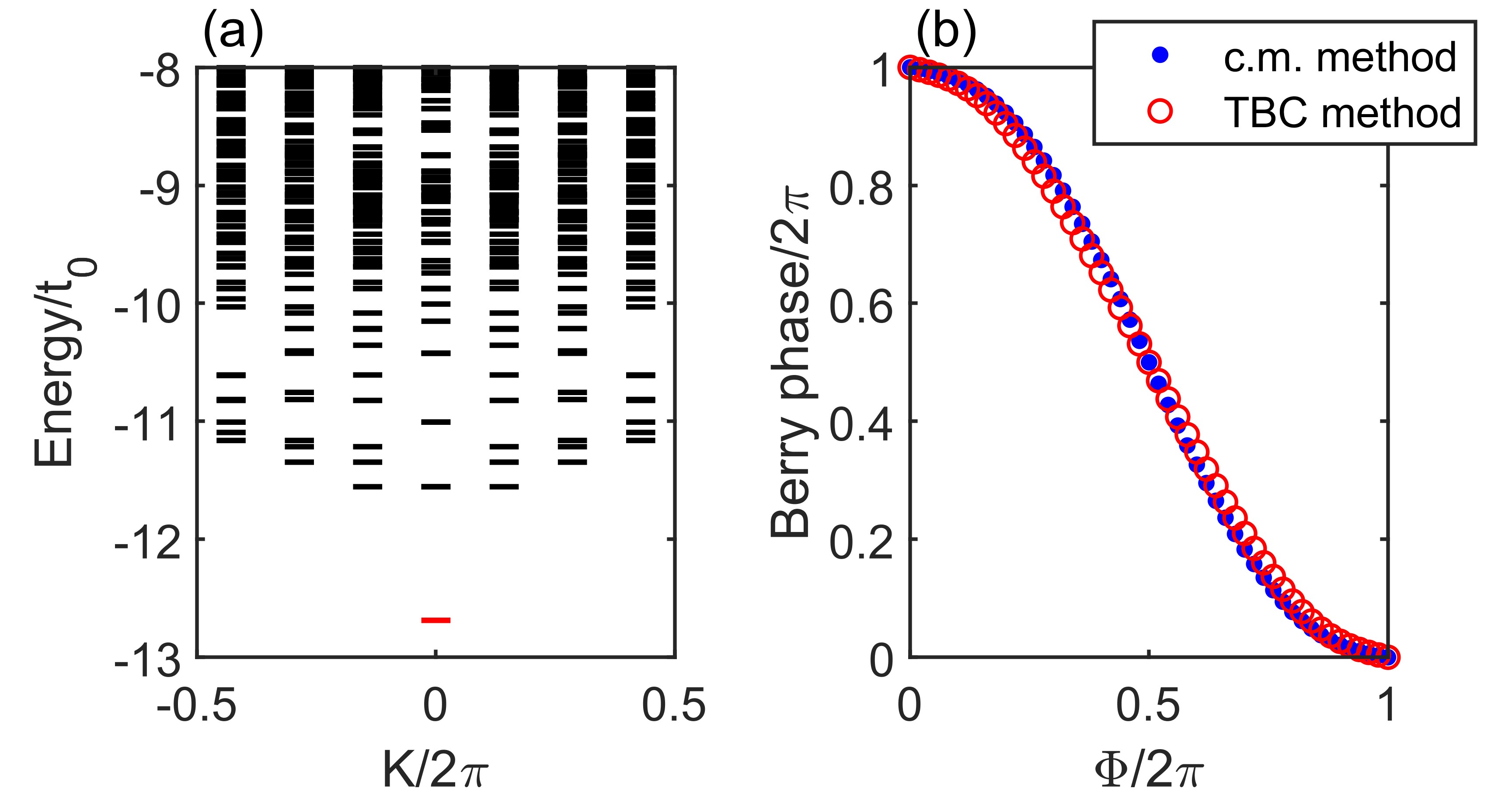}
  \caption{\label{fig:FIG_AAH_manybody_nonInteracting}
 (a) Low-lying spectrum of the non-interacting off-diagonal AAH model at $\nu=1$ filling when $\Phi = 0$.
 The unique gapped ground state is marked by red color.
 (b) Berry phase of the ground state as a function of modulation phase $\Phi$.
 Red circles are calculated through the TBC [Eq.~\eqref{eqn:ZakPhase_TBC_Definition}].
 Blue dots are calculated using the c.m. momentum states [Eq.~\eqref{eqn:ZakPhase_PBC_WilsonLoop_K}].
 We have subtracted the classical polarization for convenience.
 Parameters are chosen as $N=7$, $L=7$ (the total length of lattice is $L/b=21$), $V=0$.
 Other parameters are fixed as $b=1/3, t_0=1$ and $\lambda=0.5$.
  }
\end{figure}

\subsection{Many-body AAH model with long-range interaction}
First, let us focus on many-body ground states of the non-interacting AAH model.
When the interaction is absent, it is known that the ground state is gapped at integer filling $\nu=N/L=1$ with nonzero $t_0$, $\lambda$ for hard-core bosons.
Under the PBC, we utilize the method introduced in Sec.~\ref{Sec:sub_Cotranslation_Sym_and_CM_momentum} to construct and diagonalize c.m. Bloch Hamiltonian using the exact diagonalization method.
One can use the numerical method dubbed the seed-state algorithm introduced in Ref.~\citep{PhysRevA.95.063630} to efficiently achieve it.
The low-lying energy spectrum of the non-interacting AAH model under PBC is shown in Fig.~\ref{fig:FIG_AAH_manybody_nonInteracting} (a).
It can be seen that there is a unique and gapped ground state with c.m. momentum $K=0$.
This ground state corresponds to an insulating phase where the lowest band in the single-particle AAH model is occupied.
Next, we calculate the Berry phase of this instantaneous ground state via the TBC method [Eq.~\eqref{eqn:ZakPhase_TBC_Definition}] and the c.m. momentum method [Eq.~\eqref{eqn:ZakPhase_PBC_WilsonLoop_K}].
By applying the c.m. method, according to Eq.~\eqref{eqn:Berry_phase_CM_WilsonLoop}, the multi-particle Wilson loop for this unique ground state reads as
\begin{eqnarray}
\phi _{\rm{c.m.}} &=& {\rm{Arg}}\left( {\langle {u_{K = 0}}|{u_{K =  - 2\pi }}\rangle } \right). 
\end{eqnarray}
Numerical results are shown in Fig.~\ref{fig:FIG_AAH_manybody_nonInteracting} (b), in which both methods agree well with each other, despite some tiny differences attributed to the finite-size effect.
The Berry phase is a function of the modulation phase and continuously changes from 0 to $2\pi$.
According to Eq.~\eqref{eqn:ChernNum_1p1D_formula}, the Chern number for this adiabatic pumping process is $C=-1$, indicating a quantized shift of all particles.
It can also be confirmed that this result is consistent with the single-particle topological band theory.
\begin{figure}
  \includegraphics[width = \columnwidth ]{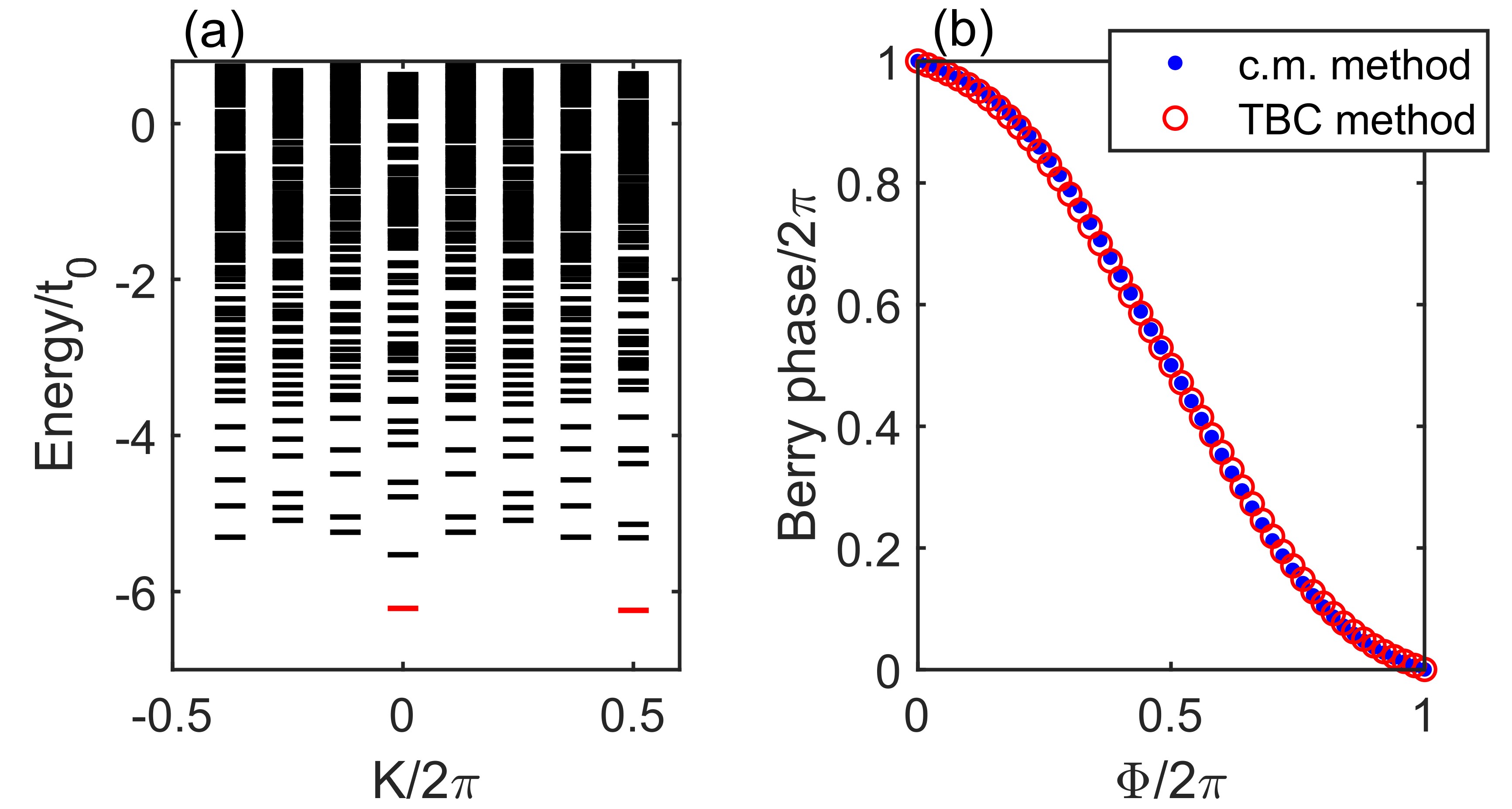}
  \caption{\label{fig:FIG_AAH_manybody_Interacting}
 (a) Low-lying spectrum of the non-interacting off-diagonal AAH model at $\nu=1/2$ filling when $\Phi = 0$.
 The two gapped ground states are marked by red color.
 (b) Berry phase of the two-fold ground states as a function of modulation phase $\Phi$.
 Red circles are calculated through the TBC [Eq.~\eqref{eqn:ZakPhase_TBC_Definition}].
 Blue dots are calculated using the c.m. momentum states [Eq.~\eqref{eqn:ZakPhase_PBC_WilsonLoop_K}].
 We have subtracted the classical polarization for convenience.
 Parameters are chosen as $N=4$, $L=8$ (the total length of lattice is $L/b=24$), $V=50$.
 Other parameters are fixed as $b=1/3, t_0=1$ and $\lambda=0.5$.
  }
\end{figure}
To further verify the relation between the two Berry phases, we introduce a long-range interaction among particles
\begin{equation}
{{\hat H}_{\rm{Int}}}(\Phi ) = {{\hat H}_{\rm{AAH}}}(\Phi )  + V\sum\limits_{i < j} {\frac{{{{\hat n}_i}{{\hat n}_j}}}{{|i - j{|^3}}}} ,
\end{equation}
which is known to support gapped ground states at fractional filling $\nu \ne 1$ \citep{PhysRevLett.110.215301, PhysRevB.86.085124, PhysRevB.88.035139}.
Here, we consider a case of $\nu=1/2$ filling.
The low-lying energy spectrum under PBC with strong interaction $|{t_0}/V| \ll 1$ is presented in Fig.~\ref{fig:FIG_AAH_manybody_Interacting} (a), where two gapped ground states appear at $K=0$ and $K=\pi$ with near-degenerate energy.
Next, we proceed to compute the Berry phases through the c.m. momentum state method and TBC method numerically, see Fig.~\ref{fig:FIG_AAH_manybody_Interacting} (b).
Note that the multi-particle Wilson loop for this two-fold ground states reads as
\begin{eqnarray}
{\phi _{{\rm{c}}.{\rm{m}}.}} &=& {\rm{Arg}}\left( {\langle {u_{K = 0}}|{u_{K =  - \pi }}\rangle \langle {u_{K =  - \pi }}|{u_{K =  - 2\pi }}\rangle } \right) .
\end{eqnarray}
It can be seen that the two methods are still in agreement.

\subsection{Few-body AAH model with nearest-neighbor interaction}

\begin{figure}
  \includegraphics[width = \columnwidth ]{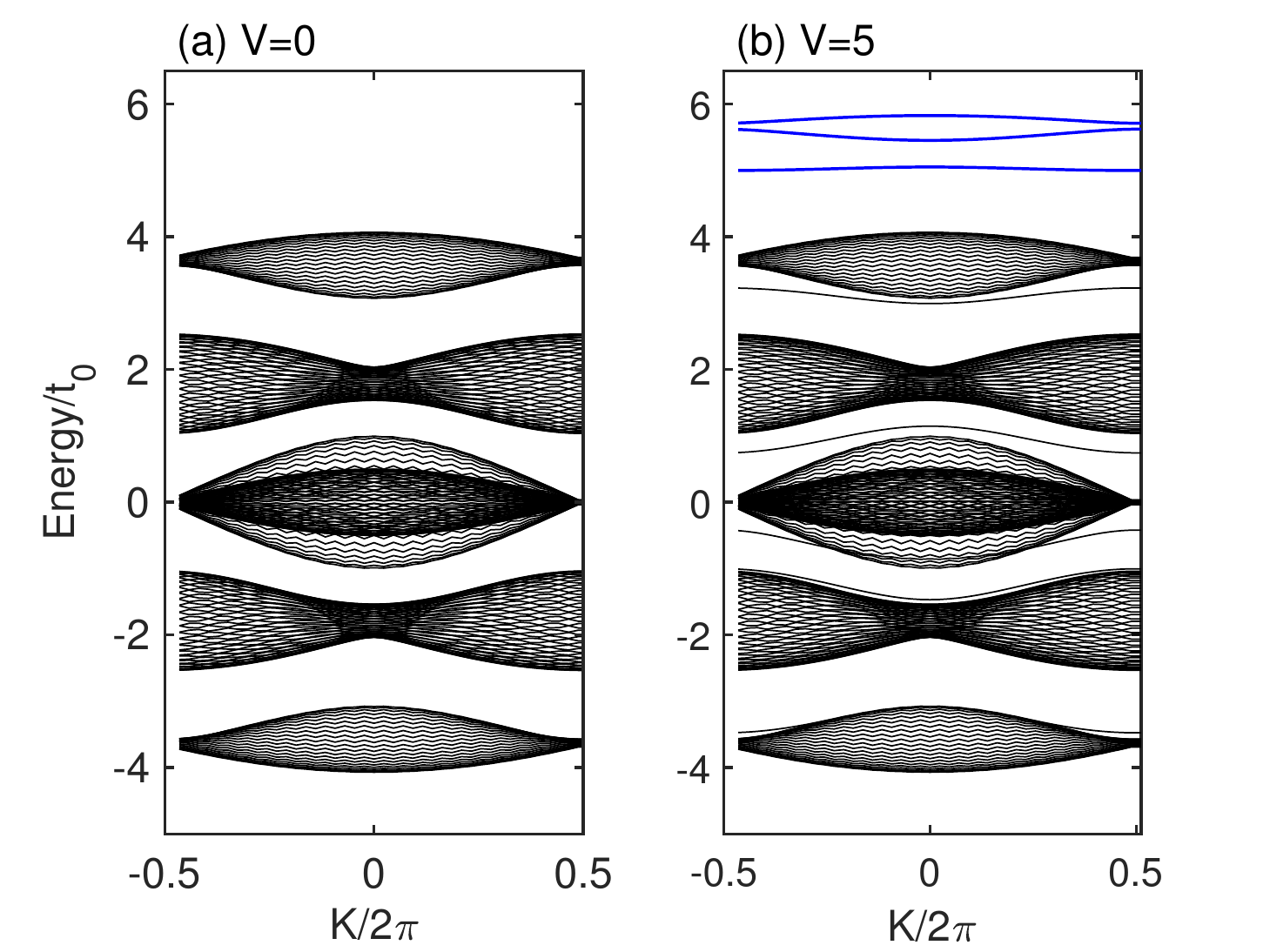}
  \caption{\label{fig:FIG_AAH_Fewbody_band}
  Instantaneous band structures of the two-particle AAH model at $\Phi=0$.
  (a-b) respectively show the spectrum under different values of interaction strength $V=0,5$.
  Isolated bands corresponding to strongly bound states are marked by blue color in (b).
  The number of cells is set to $L=43$ (the total length of lattice is $L/b=129$), and other parameters are the same as Fig.~\ref{fig:FIG_AAH_manybody_nonInteracting}.
  }
\end{figure}

Now, let us verify the relation between TBC Berry phase and the c.m. Berry phase in the few-body system with the total number of particles fixed to $N=2$.
For simplicity, let us consider a nearest-neighbor interaction between particles
\begin{equation}
{{\hat H}_{\rm{Int}}}(\Phi ) = {{\hat H}_{\rm{AAH}}}(\Phi )  + V\sum\limits_j {{{\hat n}_j}{{\hat n}_{j + 1}}} .
\end{equation}
The band structures at $\Phi=0$ with different interaction strengths are shown in Fig.~\ref{fig:FIG_AAH_Fewbody_band}.
In the absence of interaction [Fig.~\ref{fig:FIG_AAH_Fewbody_band} (a)], the spectrum is a combination of two single-particle spectra.
Because the single-particle system has three gapped bands when $b=1/3$, there are five continuum bands in the two-particle spectrum.
The continuum band corresponds to the nearly independent movement of the two particles.
From the top to the bottom, these five continuum bands correspond to five cases: (i) both particles are in the highest (single-particle) band; (ii) either of the particles is in the middle band, while the other one is in the highest band; (iii) both particles are in the middle band; (iv) either of the particles is in the middle band, while the other one is in the lowest band; (v) both particles are in the lowest band.
When the interaction strength is sufficiently strong compared to the band width, isolated bands emerge from continuum bands, see Fig.~\ref{fig:FIG_AAH_Fewbody_band} (b).
These isolated bands correspond to the bound states induced by the particle-particle interaction.
Notably, some isolated bands are well separated from the continuum bands, while others are close to the continuum band.
The upmost three isolated bands are strongly-bound states, and those isolated bands emerging between the continuum bands are weakly-bound states.

\begin{figure}
  \includegraphics[width = \columnwidth ]{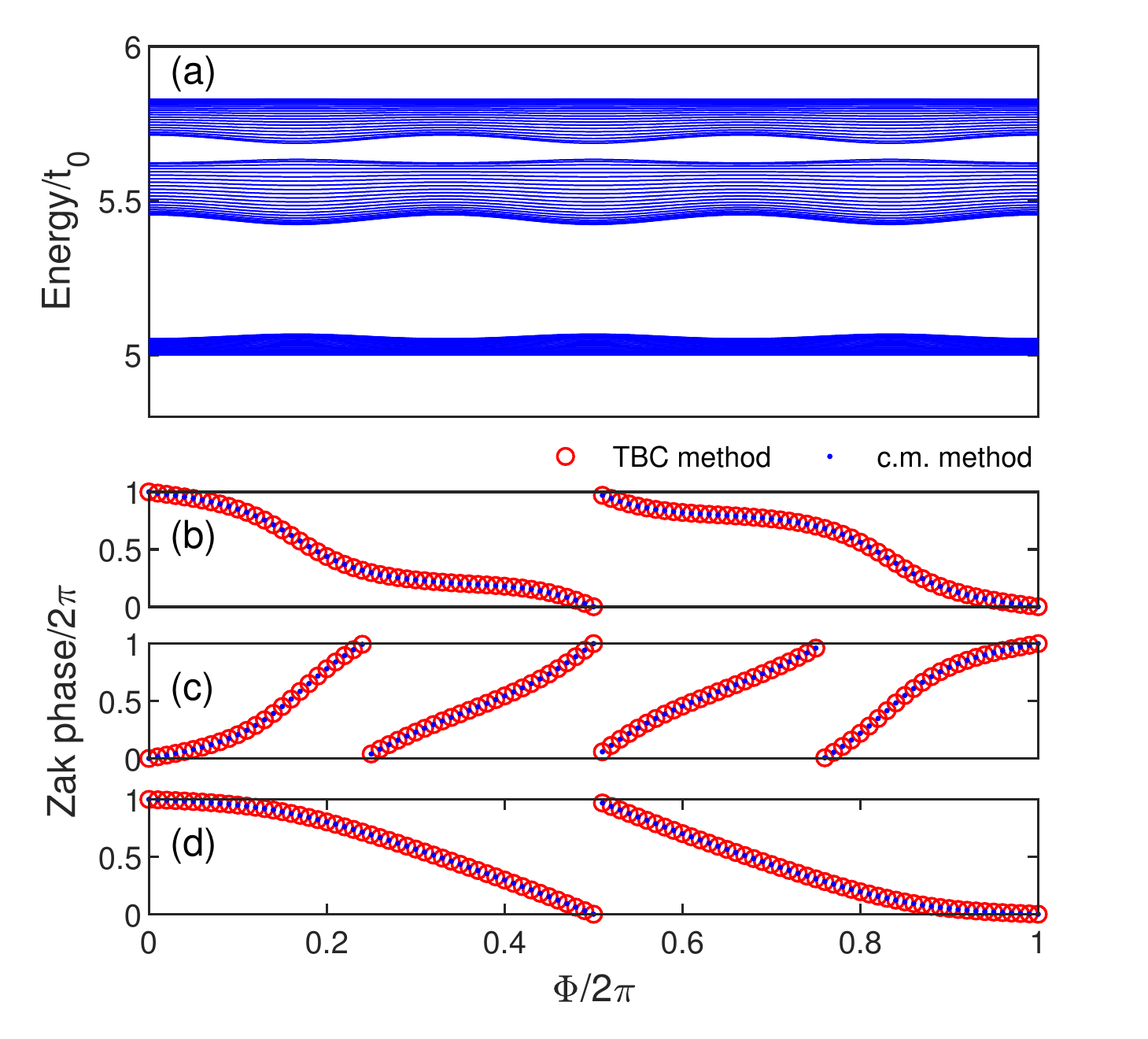}
  \caption{\label{fig:FIG_AAH_Fewbody_Pumping}
  (a) Energy spectrum of isolated bands in the two-particle AAH model as a function of the modulation phase $\Phi$.
  (b-d) Berry phases of the three gapped isolated bands (from top to bottom) as a function of the modulation parameter $\Phi$.
  Red circles are calculated through the TBC [Eq.~\eqref{eqn:ZakPhase_TBC_Definition}].
  Blue dots are calculated using the c.m. momentum states [Eq.~\eqref{eqn:ZakPhase_PBC_WilsonLoop_K}].
  Other parameters are the same as Fig.~\ref{fig:FIG_AAH_Fewbody_band} (b).
  }
\end{figure}

Next, we investigate the pumping process for the upmost three isolated bands.
For simplicity, we fix the interaction strength $V/t_0=10$.
The spectrum of the three isolated bands presented in Fig.~\ref{fig:FIG_AAH_Fewbody_band} (b) is plotted as a function of the modulation phase in Fig.~\ref{fig:FIG_AAH_Fewbody_Pumping} (a).
It can be seen that the three isolated bands stay gapped during the pumping process.
These gapped isolated bands allow us to apply the TBC method [Eq.~\eqref{eqn:ZakPhase_TBC_Definition}] and the c.m. momentum method [Eq.~\eqref{eqn:ZakPhase_PBC_WilsonLoop_K}] to calculate the instantaneous Berry phases as a function of modulation phase respectively, see Fig.~\ref{fig:FIG_AAH_Fewbody_Pumping} (b-d).
Here, the two-particle Wilson loop reads
\begin{equation}
{\cal W}_{K \to K - 4\pi }^{\left( 2 \right)} = F_K^{\left( 2 \right)}F_{K - 2\delta K}^{\left( 2 \right)} \cdots F_{K - 4\pi  + 2\delta K}^{\left( 2 \right)},
\end{equation}
where we have chosen that the particle number $N=2$ and the cell length $L=43$ are coprime.
Clearly, the TBC approach and the c.m. momentum approach are again in good agreement.
According to Eq.~\eqref{eqn:ChernNum_1p1D_formula}, the Chern number can be extracted from the winding of the Berry phase.
Therefore, from top to bottom in Fig.~\ref{fig:FIG_AAH_Fewbody_Pumping} (b-d), we can obtain the Chern number of the three cases $C=\{-2, +4, -2\}$, which are twice the values of the single-particle bands, respectively.

\section{Summary and discussions}
\label{Sec:Summary}

In this article, we have systematically studied the topological invariants defined through the TBC method and the c.m. momentum method in the presence of co-translational symmetry.
Under the TBC, one can define a TBC Berry phase through the twist angle.
Such a kind of definition is based on the modern polarization theory, where polarization is related to the adiabatic current induced by the change of lattice potential \citep{PhysRevB.27.6083, PhysRevB.47.1651,PhysRevB.48.4442, RevModPhys.66.899}.
The gauge invariance is discussed in details, and we provides a useful method to fix the gauge in practical calculations.
Notably, we have considered the non-Abelian form of the Berry phase to study the topological property of multiple gapped eigenstates.
Hence, it can be applied to non-interacting and interacting systems.
Since the twist angle is shown to adiabatic relate different eigenstates, one should involve all these related states to calculate the Berry phase instead of computing them solely.

On the other hand, we have discussed how to construct the c.m. momentum basis according to the co-translation symmetry.
To investigate the topological property of the c.m. momentum state, we introduce the multi-particle Wilson loop, which is a generalization of the single-particle version.
It allows us to define the c.m. Berry phase, and later it is shown to capture the gauge-invariant geometric phase among gapped multi-particle states.
Such kind of definition makes the c.m. Berry phase applicable for both few-body systems and many-body systems.
In addition, this method is also commensurate with the single-particle case.

It is shown that the twist angle connects different c.m. momentum sectors adiabatically.
By utilizing the perturbative nature of the twist angle, we uncover the fact that the TBC Berry phase can be equivalently formulated by c.m. momentum states.
Importantly, we prove that the TBC Berry phase is deeply related to the c.m. Berry phase obtained from the multi-particle Wilson loop.
Since the Chern number can be written as the winding of Berry phase, the Chern number defined through the TBC can be equivalently computed through the c.m. momentum state.
The use of c.m. momentum state is beneficial for numerical calculations.
We can work in the c.m. momentum subspace, which greatly reduces the dimension of the multi-particle Hilbert space.
In particular, Eq.~\eqref{eqn:BerryPhase_WilsonLoop_Manybody_simplification} suggests a method to efficiently calculate the Berry phase for many-body systems with degenerate ground states.

To verify our arguments, we apply our methods to the AAH model.
In the many-body condition, we use the TBC method and the c.m. momentum method to compute the Berry phase of the unique gapped ground state at $\nu=1$ filling.
In the few-body condition, similarly, we investigate the isolated bound-state band induced by interactions through these two methods.
In both cases, the c.m. approach is consistent with the conventional TBC method.
The numerical results show that the multi-particle Wilson loop can well capture the topological property of the many-body ground state even if there is only one state.
This is quite different from the generic single-particle Wilson loop in a non-interacting system or few-body system, in which one needs a number of states to form the loop.
With the multi-particle Wilson loop, one can avoid the integration of the twist angle and reduce the computation effort in multi-particle systems.
The equivalence between the topological invariant defined through the TBC and the c.m. momentum state is of importance.
It can be seen that the c.m. momentum states of the gapped ground state are correlated, which plays a fundamental role in formulating the topological invariant.
This offers a benefit to the understanding of multi-particle topological states.
Since the multi-particle Wilson loop formulated by the c.m. momentum states can be applied to both many-body ground states and few-body bands, it is appealing to investigate the relation between the few-body and many-body topological states in future.
Meanwhile, the emergence of topological bound states in few-body systems may have some relations to the many-body fractional topological state \citep{PhysRevLett.111.126802, PhysRevResearch.5.013112}.
It is worthwhile to investigate the nature of fractional topological states through the c.m. momentum state method in future.

\begin{acknowledgments}
This work is supported by the National Key Research and Development Program of China (Grant No. 2022YFA1404104), the National Natural Science Foundation of China (Grant No. 12025509, 11874434), and the Key-Area Research and Development Program of GuangDong Province (Grant No. 2019B030330001).
L.L. is supported by the NSFC (Grant No. 12247134).
Y.K. is partially supported by the NSFC (Grant No. 11904419, No. 12275365). 
\end{acknowledgments}

\appendix

\section{Co-translation symmetry and center-of-mass momentum for fermions}
\label{appendix:co_translation_fermions}
In this section, we demonstrates how to construct the c.m. momentum basis through the co-translation symmetry when the particle is fermionic.
In 1D, we write the $N$-fermion basis in position space as
\begin{equation}
|x_1,x_2,\cdots, x_N \rangle = \hat c_{{x_1}}^\dag \hat c_{{x_2}}^\dag  \cdots \hat c_{{x_N}}^\dag |0\rangle 
\end{equation}
in which the position of the particle is in ascending order: $x_1<x_2<\cdots <x_N$, and $\hat{c}_{x_j}^\dagger$ is the fermionic creation operator satisfying the anti-commutation relation.
Under PBC, there is $\hat{c}_{L+1}^\dagger$ = $\hat{c}_{1}^\dagger$.
When particles are translated across the boundary, we should permute the order of the creation operator to the left-most side.
For example, let us consider the co-translation of the following case:
\begin{eqnarray}
\hat T\left( {\hat c_1^\dag \hat c_2^\dag  \cdots \hat c_L^\dag } \right)|0\rangle  & =& \left( {\hat c_2^\dag \hat c_3^\dag  \cdots \hat c_{L + 1}^\dag } \right)|0\rangle \nonumber \\
& =& \left( {\hat c_2^\dag \hat c_3^\dag  \cdots \hat c_1^\dag } \right)|0\rangle .
\end{eqnarray}
To make sure the position of the particle is in ascending order, we have to permute the last creation operator to the left-most side, which yields an overall phase
\begin{equation}
\hat c_2^\dag \hat c_3^\dag  \cdots \hat c_1^\dag |0\rangle  = {e^{i\pi \sum\limits_{j = 2}^L {{{\hat n}_j}} }}\hat c_1^\dag \hat c_2^\dag  \hat c_3^\dag\cdots  |0\rangle .
\end{equation}
In other words, for even particle number $N\in 2\mathbb{Z}$ in 1D, the co-translation operation obeys a twisted boundary condition with $\theta = \pi$, which is also called the \emph{anti-periodic boundary condition}.
This can be also derived by performing the Jordan-Wigner transformation to transform the fermionic system to the hard-core bosonic system.
To take into account the quantum statistic effect of fermions in multi-particle system when constructing the c.m. momentum basis, we should modify Eq.~\eqref{eqn:Translational_Operator_Eigenstates} for even particle number.
In this case, the anti-periodic boundary condition breaks general co-translation symmetry since $[\hat{H}, \hat{T}]\ne 0$.
As demonstrated in Sec.~\ref{Sec:TBC_and_CM_momentum}, it is helpful to transform the twisted boundary condition here from the boundary gauge to the periodic gauge, and then the co-translation symmetry is restored.
The fermionic c.m. momentum basis can be thus written as
\begin{equation}
|K,\beta \rangle_{\mathrm{F}}  = \frac{1}{{\sqrt {{C_\beta }} }}\sum\limits_R {{e^{iKR}}{e^{i\frac{\pi }{L}\hat x}}|R,\beta \rangle } .
\end{equation}
On the other hand, for odd particle number, the c.m. momentum basis remains the same form as the bosonic one.
With this method, our framework on c.m. momentum state is valid for fermions, and our results on the relation between TBC and c.m. momentum in Sec.~\ref{Sec:TBC_and_CM_momentum} is still applicable.
In addition, we give a brief discussion for the 2D system.
Similarly, one can specify the order of the creation operator in a 1D manner when constructing the position basis in 2D systems.
Under the PBC, the co-translation operation for fermions leads to a complicated anti-periodic boundary condition depending on the particle distribution in the lattice.
Nevertheless, it is still possible to introduce the periodic gauge to restore the co-translation symmetry, and thus the c.m. momentum basis can be constructed in the same vein.

\section{Quasi unitarity of $\mathcal{M}$}
\label{appendix:unitarity_of_M}

Below, we show that the matrix $\mathcal{M}$ mentioned in Eq.~\eqref{eqn:Connection_A_approximation_M} is a unitary matrix in the thermodynamic limit.
According to Eq.~\eqref{eqn:M_matrix_representation}, there is
\begin{equation}
{\mathcal M} = {\boldsymbol{\Psi }^\dag}{\hat{U}_{2\pi }}{{\boldsymbol{\Psi }}^{\prime} },\;{{\mathcal M}^\dag } = {\boldsymbol{\Psi }^{\prime\dag}}\hat{U}_{2\pi }^{ - 1}{{\boldsymbol{\Psi }} },
\end{equation}
in which ${\mathbf{\Psi }}^\prime  ={\mathbf{\Psi }}  \mathcal{S} $ and $\mathcal{S}$ is an orthogonal matrix that transforms the index $\mu$ to $\mu'$ according to how the eigenstate flows after the twist angle $\theta$ changes for $2\pi$.
The vector is normalized: ${\boldsymbol{\Psi }^{\dag}}{{\boldsymbol{\Psi }} } ={\boldsymbol{\Psi }^{\prime\dag}}{{\boldsymbol{\Psi }}^{\prime} } = {I_{\mathcal N}}$, and $\mathcal N$ is the number of target states.
Meanwhile, we have ${\boldsymbol{\Psi }}{{\boldsymbol{\Psi }^{\dag}} } = {\boldsymbol{\Psi }^{\prime}}{{\boldsymbol{\Psi }}^{\prime\dag} } = {\sum\nolimits_\mu  {|{{ \psi }_\mu }\rangle \langle {{ \psi }_\mu }|} } = 1$ in the subspace spanned by target states.
There is
\begin{eqnarray}
{\mathcal M}{{\mathcal M}^\dag } &=& {\boldsymbol{\Psi }^\dag}{\hat{U}_{2\pi }}{{\boldsymbol{\Psi }}^{\prime} }{\boldsymbol{\Psi }^{\prime\dag}}\hat{U}_{2\pi }^{ - 1}{{\boldsymbol{\Psi }} } \nonumber \\
 &=& {\boldsymbol{\Psi }^\dag}{\hat{U}_{2\pi }}\left( {\sum\limits_\mu  {|{{ \psi }_\mu }\rangle \langle {{ \psi }_\mu }|} } \right)\hat{U}_{2\pi }^{ - 1}{{\boldsymbol{\Psi }} } \nonumber \\
 &=& {\boldsymbol{\Psi }^\dag}\left( {\sum\limits_\mu  {|{\psi _\mu }\left( {2\pi } \right)\rangle \langle {\psi _\mu }\left( {2\pi } \right)|} } \right){{\boldsymbol{\Psi }} },
\end{eqnarray}
where ${|{\psi _\mu }\left( {\varphi } \right)\rangle }$ is the eigenstate under periodic gauge, as already mentioned in the main text.
Using the expansion~\eqref{eqn:PBC_psi_expand} for these states, we find
\begin{eqnarray}
\sum\limits_\mu  {|{\psi _\mu }\left( {2\pi } \right)\rangle \langle {\psi _\mu }\left( {2\pi } \right)|}  &=& \sum\limits_\mu  {|{{ \psi }_\mu }\rangle \langle {{ \psi }_\mu }|}  + O\left( {\frac{1}{L}}  \right)  \nonumber \\
&=& 1  + O\left( {\frac{1}{L}}  \right) 
\end{eqnarray}
which implies that $\sum_\mu  {|{\psi _\mu }\left( {2\pi } \right)\rangle \langle {\psi _\mu }\left( {2\pi } \right)|} $ is close to the identity matrix in this subspace.
Hence, we find ${\mathcal M}{{\mathcal M}^\dag } = {I_{\mathcal N}}$ in the thermodynamic limit.
One can also prove that ${{\mathcal M}^\dag }{\mathcal M} = {I_{\mathcal N}}$ using the same analysis.
In summary, we have shown that the matrix ${\mathcal M}$ is approximately a unitary matrix in the thermodynamic limit.
Similar conclusions can be found in Refs.~\citep{Loring_2010,hastings2010almost}.

\bibliography{Bib_Equivalence}

\end{document}